\documentclass[twocolumn,pre,floats,aps,amsmath,amssymb,nofootinbib,usenatbib]{mn2e}
\usepackage{graphicx}
\usepackage{bm}
\usepackage{pgf}
\usepackage[T1]{fontenc}
\usepackage{natbib}
\usepackage{aas_macros}
\usepackage[english]{babel}
\usepackage{tabu}
\usepackage{subcaption}
\usepackage{amsmath}
\usepackage{amssymb}
\usepackage{mathrsfs}
\usepackage{hyperref}

\let\pgfimageWithoutPath\pgfimage 
\renewcommand{\pgfimage}[2][]{\pgfimageWithoutPath[#1]{pictures/#2}}

\graphicspath{{./pictures/}}

\newcommand{\revised}[1]{#1}
\newcommand{\revisedvtwo}[1]{#1}

\begin{document}

\date{Accepted 2018 July 16. Received 2018 July 16; in original form 2018 February 21}

\pagerange{\pageref{firstpage}--\pageref{lastpage}} \pubyear{2018}

\citestyle{egu}
\bibliographystyle{mn2e}

\title[Vortex Survival in VSI]{Vortex Formation and Survival in Protoplanetary Disks subject to Vertical Shear Instability}
\author[N.Manger and H. Klahr]
{Natascha Manger$^{1}$\thanks{E-mail:manger@mpia.de} and Hubert Klahr$^{1}$
\\
$^{1}$Max Planck Institute for Astronomy, K{\"o}nigstuhl 17, 69117 Heidelberg, Germany}

\maketitle

\label{firstpage}

\begin{abstract}
\revised{
Several protoplanetary disks observed by ALMA show dust concentrations consistent with particle trapping in giant vortices. The formation and survival of vortices is of major importance for planet formation, because vortices act as particle traps and are therefore preferred locations of planetesimal formation. 
Recent studies showed that the vertical shear instability (VSI) is capable of generating turbulence and small vortices in protoplanetary disks that have the proper radial and vertical stratification and thermally relax on sufficiently short time scales. But the effect of the azimuthal extend of the disk is often neglected as the disks azimuth is limited to $\Delta \phi \leq \pi/2$.
We aim to investigate the influence of the azimuthal extent of the disk on the long-term evolution of a protoplanetary disk and the possibility of large vortices forming. 
To this end, we perform 3-dimensional simulations for up to 1000 local orbits using different values of $\Delta \phi = \pi/2 $ to $2\pi$ for VSI in disks with a prescribed radial density and temperature gradient cooling on short timescales.
We find the VSI capable of forming large vortices which can exist at least several hundred orbits in simulations covering a disk with $\Delta \phi \geq \pi$. This suggests the VSI to be capable to form vortices or at least to trigger vortex formation via a secondary instability, e.g. Rossby Wave Instability or Kelvin Helmholtz Instability.
}
\end{abstract}


\section{Introduction}
\label{sec:intro}

Turbulence in disks around young stars is still one of the most interesting questions in modern astrophysics \citep{Turner+2014}. \cite{Balbus+Hawley1991} introduced the magneto-rotational instability (MRI)\revised{ as a promising source of turbulence with an alpha viscosity \citep{Shakura+Sunyaev1973} large enough to explain angular momentum transport on timescales set by observations}. But more recent work shows the MRI to be hampered by non-ideal magnetic effects such as resistivity or ambipolar diffusion \citep{Lesur+2014}. They show that the instability can be damped efficiently in parts of the disk by low ionization \revised{fractions}, where then other sources of turbulence can and have to be considered \citep{Lyra+Klahr2011}.

Several possible mechanisms for pure hydrodynamic turbulence have since been proposed, acknowledging the fact that \revised{a Keplerian flow with radially} increasing angular momentum profile \revised{is} hydrodynamic stable as can be seen in a most general way from the Solberg-Hoiland criteria, which is derived for no thermal relaxation \citep{Ruediger+2002}. The global baroclinic instability (aka Subcritical Baroclinic Instability) \citep{Klahr+Bodenheimer2003,Petersen+2007a,Petersen+2007b, Lesur+Papaloizou2010} and convective overstability (COS) \citep{Klahr+Hubbard2014,Lyra2014} rely on radial temperature and density stratifications introducing convective cells in disks with moderate cooling times \revised{ on the order of $\tau = 1/(\gamma \Omega)$}. The Vertical Shear Instability (VSI) \citep{Nelson+2013,Richard+2016,Stoll+Kley2014} relies on short cooling times \revised{to remove the stable vertical stratification} to tap into the energy of the vertical shear. 

In all of those turbulence models, the formation of vortices has been observed. \cite{Raettig+2013} showed them in local simulations of subcritical baroclinic instability, \citep{Flock+2015} in a global context at the edge of the MRI dead zones. The Rossby Wave Instability \citep{Papaloizou+Pringle1984,Papaloizou+Pringle1985,Lovelace+1999,Li+2000,Li+2001} has been shown to break axisymmetric rings into large vortices. \cite{Richard+2016} showed the possibility of small vortex formation within disks susceptible to VSI \revised{and \cite{Latter+Papaloizou2018} found the Kelvin Helmholtz Instability acting as a parasitic instability on the VSI modes and forming small vortices in the disk}. 

Because vortices are naturally identified with local pressure maxima in the context of PPDs, they act as particle traps \citep{Barge+Sommeria1995}. 
They are found to enhance the dust to gas ratio locally, aiding planetesimal formation via triggering the streaming instability \citep{Raettig+2015} and accelerate core growth for giant planets via pebble accretion \citep{Klahr+Bodenheimer2006}. Large vortices have also been discussed as explanation for features found in (sub-)mm observations of protoplanetary disks with ALMA \citep{vanDerMarel+2013} and VLA \citep{Carrasco-G.+2016}.

In this work we investigate vortex formation and survival in protoplanetary disk simulations undergoing Vertical Shear Instability, reexamining the work of \cite{Richard+2016}. We find that once we loosened the restriction to small azimuthal domains (large azimuthal wavenumbers m) and allowed for simulation domains of $\varphi = 180^\circ$ and $\varphi = 360^\circ$, we find large vortices forming in the disks after a few hundred orbits. We also find these larger vortices to survive for hundreds of local orbits, making them excellent particle traps and candidates for planetesimal formation sites. With this, we stress again that non-axisymmetry plays an important role in assessing disk turbulence features \citep{Klahr+1999}.

In section \ref{sec:theory}, we will shortly revisit the theoretical background to our simulations. Section \ref{sec:setup} lays out the numerical set-up used in all computations. In Section \ref{sec:results} we present our results and in section \ref{sec:discussion}, we discuss them in context of recent literature. Finally, section \ref{sec:summary} summarizes our findings and presents an outlook on future work.

\section{Theoretical background}
\label{sec:theory}

In this work, three instabilities have to be considered. We implement our simulation with a steady-state model susceptible to the growth of the Vertical Shear Instability, which we identify for this paper as the main driver of turbulence in the short cooling time regime. The Rossby-Wave-Instability is important in the subsequent formation of giant vortices in the disk. The elliptic instability then has to be considered acting inside the formed vortices as a driver of turbulent vortex substructures and possible vortex destruction.

\subsection{Vertical Shear Instability (VSI)}

Protoplanetary disks with radial gradients in temperature and entropy do not rotate on cylinders but have an angular frequency depending on height $\Omega =\Omega(R,z)$ and therefore exhibit vertical shear and a non-zero vertical epicyclic frequency, i.e.\ $\kappa_z^2 = \frac{\partial_z}{R^2}\Omega^2 R^4 \ne 0$. The stability of these disks against axisymmetric, adiabatic perturbations can be in general described through the Solberg-Hoiland criteria \citep{Ruediger+2002}, \revised{which in general predict stability due to the vertical stratification of the disk}. For nearly locally isothermal disks,\revised{ i.e. disks with a short cooling time, or vertically adiabatic disks \citep{Pfeil+Klahr2018}} however, the stabilizing effect of vertical buoyancy against axisymmetric perturbations is diminished. 
This leads to the Vertical Shear Instability, first described for radiative zones of differentially rotating stars by \cite{Goldreich+Schubert1967} and \cite{Fricke1968}. 
The instability has been more recently suggested to also operate in protoplanetary disks by \cite{Urpin2003} and shown by \cite{Nelson+2013}. In the limit of thin, locally isothermal disks $((H/R)\ll 1)$, instability is determined by the criterion:

\begin{equation}
\frac{\partial j^2}{\partial R^2} - \frac{k_R}{k_Z}\,\frac{\partial j^2}{\partial Z^2} < 0
\end{equation}

\revised{with $j$ the specific angular momentum of the disk and $k_R$ and $k_Z$ the radial and vertical wavenumber. }Instability then occurs for nearly vertical modes with $k_R/k_Z>R/H$ tapping into the free energy provided by the vertical shear \citep{Nelson+2013}. 

\cite{Nelson+2013,Stoll+Kley2014,Stoll+Kley2016} and \cite{Flock+2017} showed turbulent stresses for VSI active disks to be in the range $\alpha \approx 5\cdot 10^{-5} - 5\cdot 10^{-4}$.
\cite{Richard+2016} reported small vortex formation to be possible for certain disk parameters.

\subsection{Rossby Wave Instability (RWI)}

The Rossby Wave Instability, first described in \cite{Lovelace+1999,Li+2000} and \cite{Li+2001}, is a global instability occuring in rotating flows that exhibit a local extremum in pressure or vortensity. Such an extremum can be archived e.g. at the inner edge of a dead zone \cite{Lyra+MacLow2012}. The criterion for instability is an extremum in the function:
\begin{equation}
\mathscr{L}= \frac{\Sigma}{\omega_Z}\,\left(\frac{P}{\Sigma^\gamma}\right)^{2/\gamma} \qquad ,
\label{eqn:RWI}
\end{equation}
\revised{with $\Sigma$ denoting the column density, $\omega_z = (\bm{\nabla}\times \bm{v})_Z$ the z-component of vorticity, $P$ the vertically integrated pressure and $\gamma$ the adiabatic index.}

In a region around the extremum, Rossby waves are trapped in a standing pattern amplifying with time, which can be observed as a vortex at the extremum. Outside, the Rossby waves emit density waves due to gradients in vorticity \citep{Meheut+2010}.

\subsection{Elliptic instability}

The elliptic instability is a linear parametric instability. It destabilizes elliptic streamlines once the vortex turnover time matches one or more inertial frequencies of the underlying flow field by creating a positive resonance. 
\cite{Lesur+Papaloizou2009} studied the instability in the context of accretion disks, showing the instability to be most effective for vortex aspect rations $\chi$=semi-mayor axis/semi-minor axis to be in the range of $ 1 < \chi < 4 $ for 3 dimensional disks. A weaker second range for the instability occurs for $\chi > 6$, the region in between is stable for disks without vertical stratification. When vertical stratification is introduced, vortices become unstable for all $\chi$, the growth rate however depends on the strength of the stratification for vortices with aspect ratios $\chi > 4$. Vortices with smaller aspect ratios are always highly unstable independent of the strength of the stratification.


\section{Simulation Setup}
\label{sec:setup}
We conduct 3 dimensional simulations using the magneto-hydrodynamics code PLUTO\footnote{http://plutocode.ph.unito.it}. In this work, we use 2 coordinate systems. The simulations are carried out on a spherical grid $(r,\theta,\varphi)$, the model setup and analysis are presented in cylindrical coordinates $(R,\phi,Z)$,

We implement the disk setup following \cite{Nelson+2013}. The gas density is defined by
\begin{equation}
\rho = \rho_0 \left(\frac{R}{R_0}\right)^p \exp\left(-\frac{Z^2}{2\,H^2}\right)
\end{equation}
with the disk scale height $H =\frac{c_s}{\Omega_K}$. We chose a ideal equation of state $\rho e = \frac{P}{\gamma -1}$ with the specific internal energy $e$,the adiabatic index $\gamma = 5/3$\revisedvtwo{\footnote{The value for $\gamma$ in protoplanetary disks is closer approximated by 1.44 due to the diatomic nature of molecular hydrogen. We instead use the standard $\gamma =5/3$ of PLUTO. This does not significantly affect the results due to the fast cooling applied in this work.}} and $P = c_\mathrm{s}^2 \rho$ using a radially changing \revised{isothermal} sound speed $c_\mathrm{s}^2 = c_0^2 \left(\frac{R}{R_0}\right)^q $. Note that the temperature is related to the isothermal sound speed via $c_\mathrm{s}^2 = kT/\mu m_H $ with $k$ denoting the Boltzmann constant, $\mu$ the mean molecular weight and $m_H$ the atomic mass of hydrogen, thus defining a radial temperature gradient in the disk.
 
For all our simulations, we choose $-\frac{2}{3}$ and $-1$ for $p$ and $q$ respectively, satisfying the requirements set for the VSI.

The $q=-1$ is close to the maximum of $q=-1.5$ that one can expect in a viscously heated disk (see Eq. 11 in \cite{Belletal1997}) in regions which are
dominated by icy grains. The value for $p=-\frac{2}{3}$ is chosen to be consistent to models in which we try to simulate the COS, by using an identical setup as
described here, but using a longer cooling time. This particular value of $p$ has no significant impact on the VSI, but helps to maximise the radial buoyancy for the given $p$ expressed in the radial 
Brunt-Vaissala frequency $N_r^2$ \citep{Klahr+Hubbard2014}.

The geometrical scale height is then constant throughout the disk with $H/R = c_0/v_{\rm Kepler} = 0.1$, which is also nice to evaluate the simulation. 
$H/R$ might actually be smaller in a real circumstellar disk \citep{Belletal1997} and we tested a value of $H/R = 0.05$, which was 
also the value choice in the work by \cite{Stoll+Kley2014}, yet \citep{Flock+2017} again uses values of $H/R = 0.1$ for the outer disk. 
$H/R = 0.05$ simulations are computationally more expensive, as the pressure scale height has to be resolved by at least
as many cells as in the $H/R = 0.1$ case, leading to $8$ times more cells and reducing the time step by a factor of two.
Once our $H/R = 0.05$ simulations are finished, we will publish them in comparison to the $H/R = 0.1$ cases, but \revised{we claim that smaller values of $H/R$ will show similar results for runs with comparable resolution per scale height}.

The scale height can be expressed as:
\begin{equation}
H \propto \left(\frac{R}{R_0}\right)^{(q+3)/2}\qquad ,
\end{equation}
and the initial angular velocity of the disk is given by
\begin{equation}
v_\phi = \Omega_\mathrm{K}\,R \left[1+q-\frac{q\,R}{\sqrt{R^2+Z^2}}+(p+q)\left(\frac{H}{R}\right)^2\right]^{\frac{1}{2}}
\end{equation}
whereas the radial and vertical velocities are set to zero. All velocity components are initially seeded with a white noise perturbation of $10^{-4}\,c_\mathrm{s}$. 

\begin{table}
\centering
\caption{Simulation parameters for all simulation conducted in this study. We give grid parameters, phi extent and resulting time and space averaged alpha values.}
\begin{tabu}{X[1,l] >{$\displaystyle}X[2,c]<{$} >{$\displaystyle}X[1,r]<{$} >{$\displaystyle}X[1,c]<{$}} \hline
Run & N_r \times N_\theta \times N_\phi & \phi_{\mathrm{max}} [^\circ] & \alpha \\\hline
p45 & 256 \times 128 \times  96 &  45 &  1.5\cdot 10^{-3}\\
p90 & 256 \times 128 \times 192 &  90 &  1.4\cdot 10^{-3}\\
p180& 256 \times 128 \times 384 & 180 &  1.2\cdot 10^{-3}\\
p360& 256 \times 128 \times 768 & 360 &  1.0\cdot 10^{-3}\\\hline
\end{tabu}
\label{tab:runs}
\end{table}
The cooling in our model is described by
\begin{equation}
\frac{dP}{dt} = -\frac{P-\rho T_\mathrm{init}}{\tau_\mathrm{relax}}
\end{equation}
where \revisedvtwo{$T_\mathrm{init}(R)$ is the initial Temperature profile and} $\tau_\mathrm{relax}$ is the relaxation time scale. \revisedvtwo{It is set to $\tau_\mathrm{relax} = dt$ and is about $2\cdot 10^{-3} /\Omega_0$ in all simulations}. This \revisedvtwo{is} 
the shortest cooling time we can realise in our explicit cooling scheme, effectively
\revisedvtwo{leading to an almost locally isothermal disk}.

In our numerical computations we use the hllc method \citep{Toro2009} with a ppm reconstruction scheme \citep{Mignone2014} for spatial integration and a 3rd order Runge-Kutta method for time integration. The mesh extends from $(0.5 r_0, \pi/2 -0.35, 0)$ to $(2 r_0,\pi/2 +0.35, \phi_\mathrm{max} )$ with $r_0$ being the radius at $r = 1$ and $\phi_\mathrm{max}$ given in table \ref{tab:runs}. We use a logarithmic grid in radial direction and a uniform grid otherwise to preserve the aspect ratio of the individual grid cells. This gives us a resolution of 18 cells per H in vertical \revised{and radial direction} and 12 cells per H in azimuthal direction. \revised{For the largest run (p360), we used about 1.4 million cpu-hours.}

We employ outflow boundary conditions in radial direction, reflective boundaries in vertical direction and periodic boundaries in azimuthal direction. To minimise mass loss in radial direction and generally wave reflection at the boundaries, we add damping layers at the inner and outer boundaries in the radial and \revised{polar} direction with $\Delta R = 1.0\,H$ and $\Delta \theta = 0.05$. Inside the damping layers, we damp the velocities to their initial values if the velocity component normal to the boundary points inside the domain. For the damping we use:
\begin{equation}
\frac{dv_x}{dt} = -\frac{v_x-v_{x,0}}{\tau_\mathrm{damp}}\cdot f^2
\end{equation}
with the damping time $\tau_\mathrm{damp} = 0.1\frac{2\pi}{\Omega}$ and $f = \frac{R-R_\mathrm{b}}{\Delta R}$ in radial and $f = \max(\frac{R-R_\mathrm{b}}{\Delta R},\frac{\theta-\theta_\mathrm{b}}{\Delta \theta})$ in meridional direction . $R_\mathrm{b}$ and $\theta_\mathrm{b}$ denote the position of the boundary of the damping layer inside the simulation domain.

\section{Results}
\label{sec:results}

We performed simulations with azimuthal extents of 45, 90, 180 and 360 degrees to determine a minimum azimuthal range on which similar behaviour as in a complete 360 degree disk can be expected. The low resolution chosen for all simulations enables simulation times close to 1000 orbits (which for $H/R = 0.05$ needs 16 times the computation cost) enables us to determine the lifetime of the formed vortices and possible influences on disk evolution an planetesimal formation. The simulation parameters are summarized in Table \ref{tab:runs}.
In this section, we present the results of our simulations, focusing first on the flow properties and later on vorticity.

\subsection{Transport properties}
We analyse the transport properties of the VSI for all runs in a subdomain of the simulation grid. The subdomain is defined as  $R=[0.7-1.8]$, $Z=[\pm 2.5 H]$, $\phi=[0,\phi_\mathrm{max}]$. This is done to avoid possible influence of the imposed boundary conditions on the results. 

\begin{figure}
\centering
\includegraphics[width=\columnwidth]{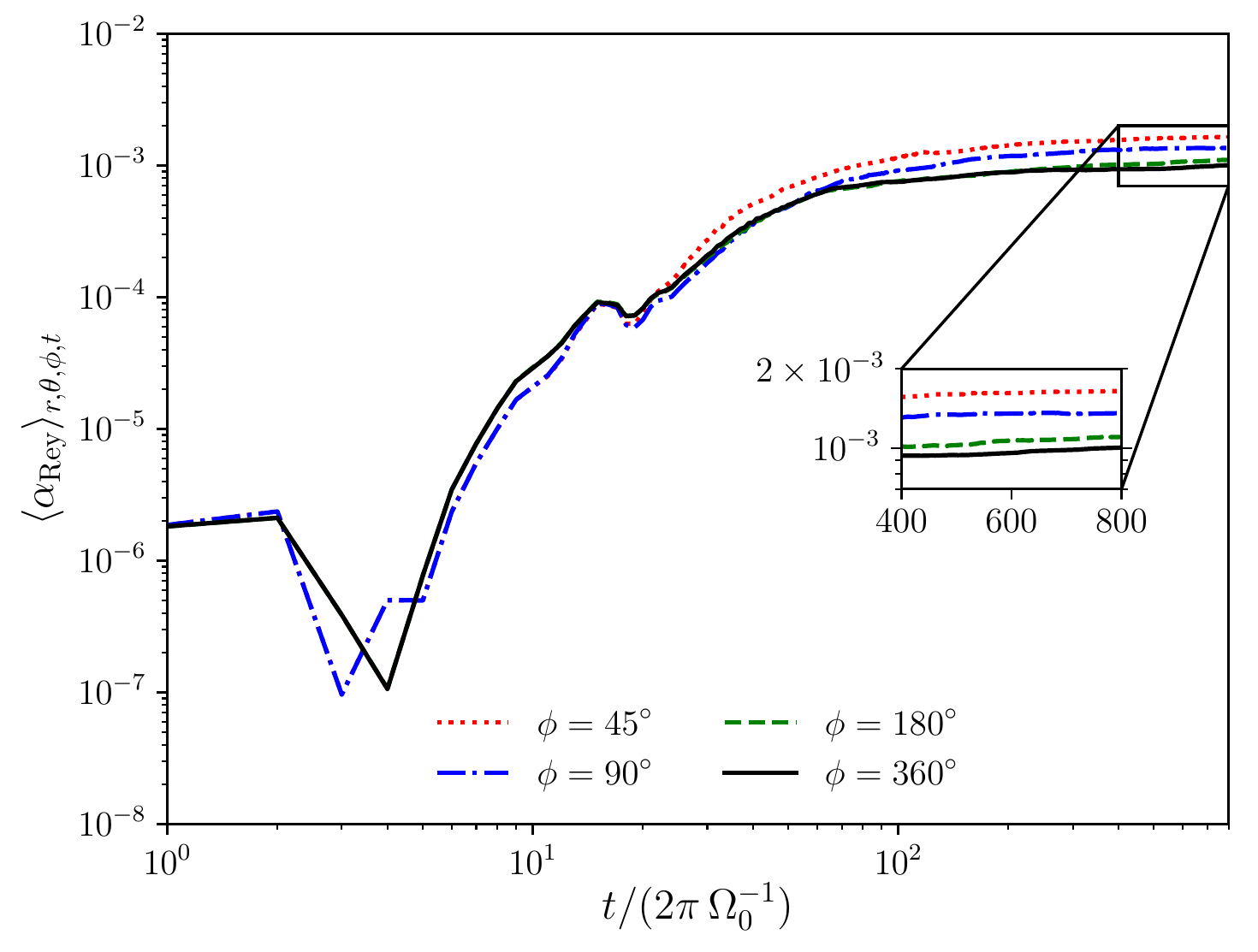}
\caption{Evolution of the alpha values over the simulation time. The alpha values are averaged over the analysis domain an as running average in time. We observe the turbulence to saturate rapidly within around 100 orbits to alpha values of around $10^{-3}$. The alpha values for the simulation with smaller azimuthal domain are slightly larger than for the larger domain. The inset highlights the evolution for the last 200 orbits.}
\label{fig:alphavTime}
\end{figure}

We look at the Reynolds stresses generated in the disk following the prescription by \cite{Klahr+Bodenheimer2003}
\begin{equation}
T_{r,\phi} = \langle \rho v_r v_\phi \rangle_{\phi,t} - \langle \rho v_r \rangle_{\phi,t} \langle v_\phi \rangle_{\phi,t}  
\end{equation}
and compute the $\alpha$-parameter of the disk  
\begin{equation}
\alpha_r (z) = \frac{\langle T_{r,\phi}\rangle_r}{\langle P \rangle_r}
\end{equation}
with $P= \langle \rho c_\mathrm{s}^2 \rangle_{\phi,t}$. This guarantees a mass weighted $\alpha$ and filters out angular momentum flux associated with mean mass transport $\langle \rho v_r \rangle$.
We plot the evolution of $\alpha$ over time \revised{in figure \ref{fig:alphavTime}}. The values are averaged over the whole analysis domain in each direction and up to the given point in time.
We find $\alpha$-values of approximately $10^{-3}$ in agreement with \cite{Stoll+Kley2014} and \citep{Nelson+2013}, but significantly higher than those reported by \cite{Richard+2016}. 
The equilibrium alpha values (table \ref{tab:runs}) can be seen to decrease with increasing azimuthal extent. This effect is caused by the use of periodic boundary conditions in phi direction, leading to a large pitch angle for the tightly wound spiral pattern for smaller phi extents.

\begin{figure}
\centering
\includegraphics[width=\columnwidth]{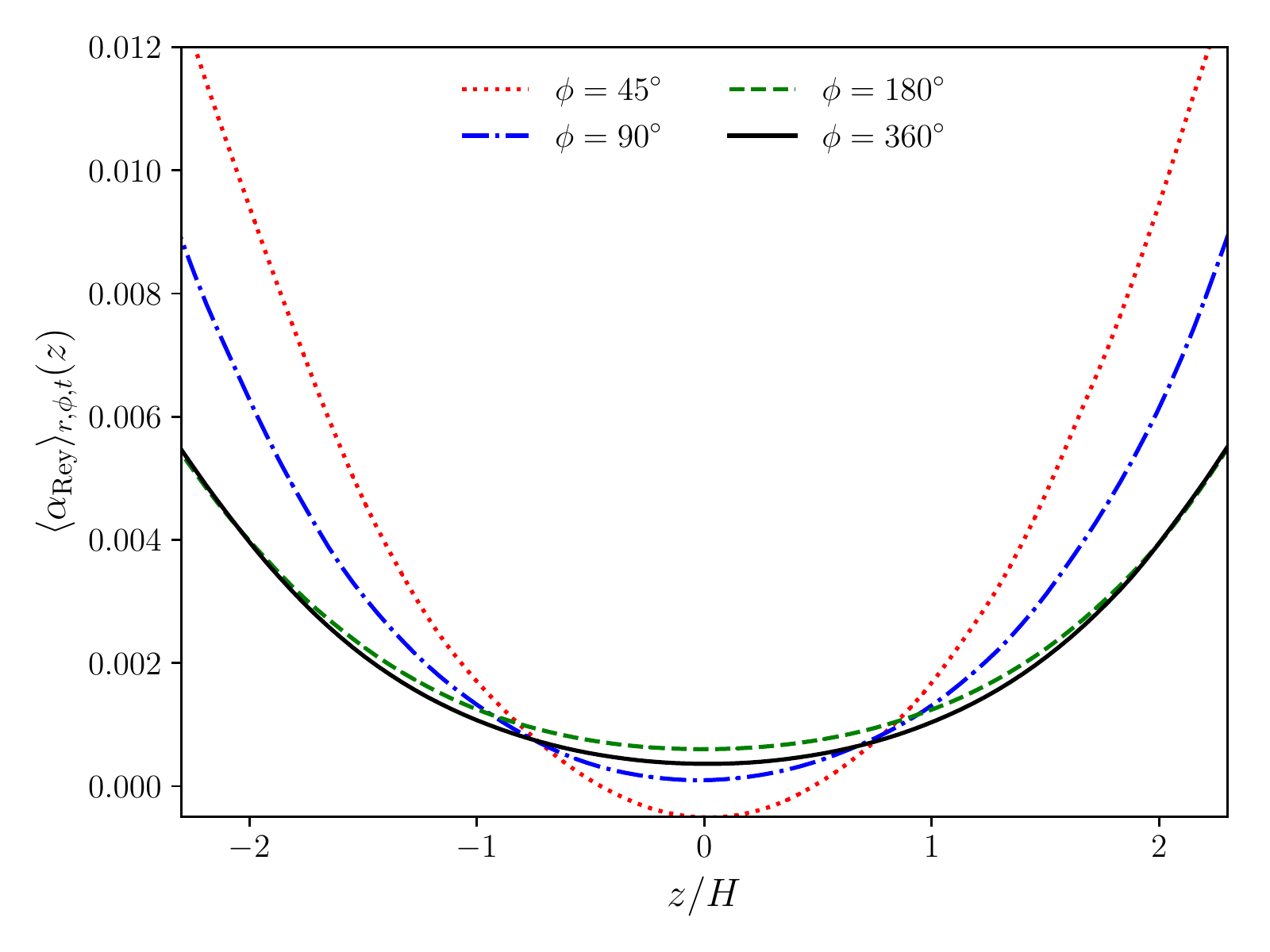}
\caption{Meridional profile of the alpha value. The alpha values are averaged over the radial and azimuthal subdomain and 500-700 orbits. We observe a steeper meridional profile and negative midplane alpha values for the simulations with small azimuthal extent. The profile for the simulations with large azimuthal extent show strictly positive alpha values.}
\label{fig:alphavR}
\end{figure}

\begin{figure}
\centering
\includegraphics[width=\columnwidth]{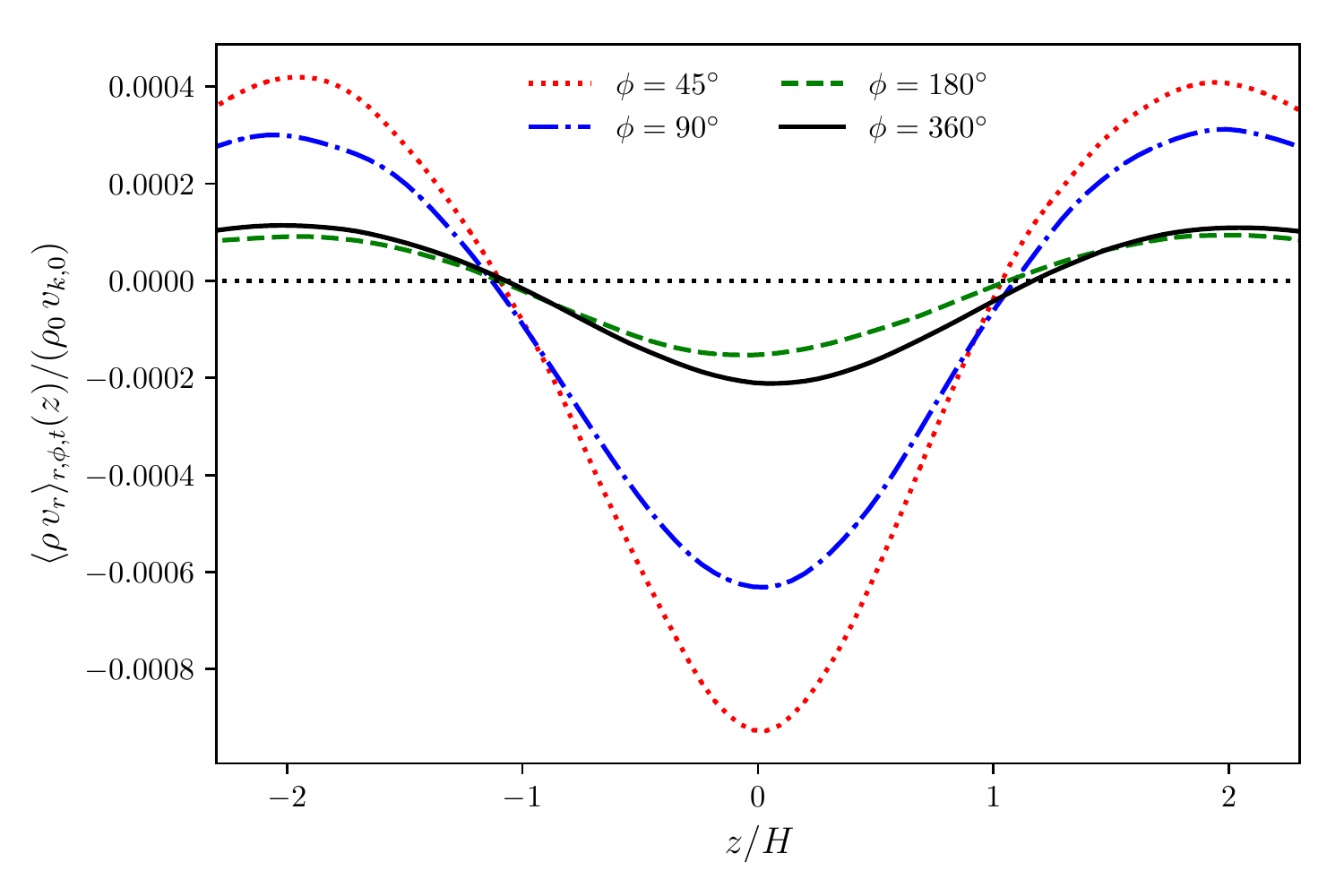}
\caption{Vertical profile of the radial mass flow. We find inward mass flow in the midplane of the disk for all simulations.}
\label{fig:rhoVrMer}
\end{figure}

To look at the vertical structure of the disk, figure \ref{fig:alphavR} compares the meridional $\alpha$ profiles of the simulated disks averaged between 500 and 700 orbits and over the radial and azimuthal subdomain. We observe different profiles for each simulation. We find the steepest profile for the simulation p45, showing high alpha values at the disk surface and negative values at the midplane, \revised{which suggest moderate inward angular momentum transport in the midplane of the disk and outward transport in the surface layer.} The profile flattens for the case p90, with positive alpha values throughout the disk and lower alpha values at the surface compared to run p45. This trend continues for the cases p180 and p360, though the latter show to be in good agreement with each other. \revised{This also suggest angular momentum is transported outward at all heights, albeit with stronger transport in the upper layers of the disk. The vertical profile of the Reynolds-stresses of the disk shows the largest angular momentum transport to occur at $z\approx\,\pm 1.6 H$ for simulation p45 and decreasing to $z \approx\,\pm 1.3 H$ for the full $2\pi$ disk.}

\revised{To \revisedvtwo{investigate this behaviour further}, we calculate the radial mass flow of the disk $\rho\, v_r$ as a function of height above the midplane in all our simulations, shown in figure \ref{fig:rhoVrMer}. We again average over the radial and azimuthal domain of the simulation and over 200 snapshots taken between 500 and 700 local orbits. We find radial mass inflow for all models in the midplane, which aligns with the findings of \cite{Stoll+Kley2017}, who find the same flow reversal applying an anisotropic viscosity model with a heightened z-viscosity component. Therefore, although there is only small to no outward \revisedvtwo{radial} angular momentum transport present in the midplane from the VSI, mass can be accreted efficiently. \revisedvtwo{The angular momentum however is transported vertically from the midplane to the upper layers of the disk, where it is then transported radially outward.} This mechanism also gives an explanation to the observed shallower vertical profiles for the models p180 and p360. We argue that due to the large azimuthal extent of the disk, the anisotropy manifests in different magnitudes and therefore leads to an overall shallower profile.} 
This supports our view that treating the disk as too high m leads to incomplete results.

\revised{
To 
prove that the vertically averaged angular momentum transport (i.e.\ our measured mean $\alpha$-value) is nevertheless 
sufficient to prescribe the mean radial mass-accretion, we calculate the average radial velocity given from steady-state viscous accretion theory and compare this to the simulation values. Integrating the vertically-averaged steady-state angular momentum equation one obtains
\begin{equation}
 \Sigma v_R = \Sigma \nu \frac{\mathrm{d}\Omega}{\mathrm{d} R} = -\frac{3}{2} \alpha \left(\frac{H}{R}\right)^2 \Sigma v_\mathrm{k} 
\end{equation}
with $\Sigma$ the disk column density, $\nu = \alpha H^2 \Omega$ the viscosity and $\mathrm{d}\Omega/\mathrm{d} R = -3/2 \Omega$. For the case p360 with $\alpha = 10^{-3}$ we get $\Sigma v_R /(\Sigma v_k) = -1.5\cdot 10^{-5}$. From the corresponding simulation run we get from integrating over figure \ref{fig:rhoVrMer} an average value of $\Sigma v_R /(\Sigma v_k) = -2.4\cdot 10^{-5}$, which agrees well to the predicted value. This shows that angular momentum transport driven by VSI
is well described in the picture of $\alpha$-viscosity, even if transport is most likely realised by travelling spiral waves rather than local Kolmogorov-like 3D turbulence.\revisedvtwo{ This also supports our claim that angular momentum is transported mainly vertically from the midplane to higher layers and then transported outward, as the vertically averaged angular momentum matches the value needed for the occurring mass transport despite the low $\alpha$ values measured in the midplane of the disk.}
Whether the turbulence is truly local, i.e.\ the dissipation of kinetic occurs also proportional to the measured $\alpha$-viscosity, was not possible to be determined in our simulations, but
should be goal for future setups. This would need the monitoring of  local heating and cooling.
}

We also look at the time evolution of the rms-velocities in the disk, defined as
\begin{equation}
v_\mathrm{rms} = \left(\frac{1}{V} \int_{V} (v_R^2 + v_Z^2) dV\right)^{0.5}  \qquad .
\end{equation}

\begin{figure}
\centering
\includegraphics[width=\columnwidth]{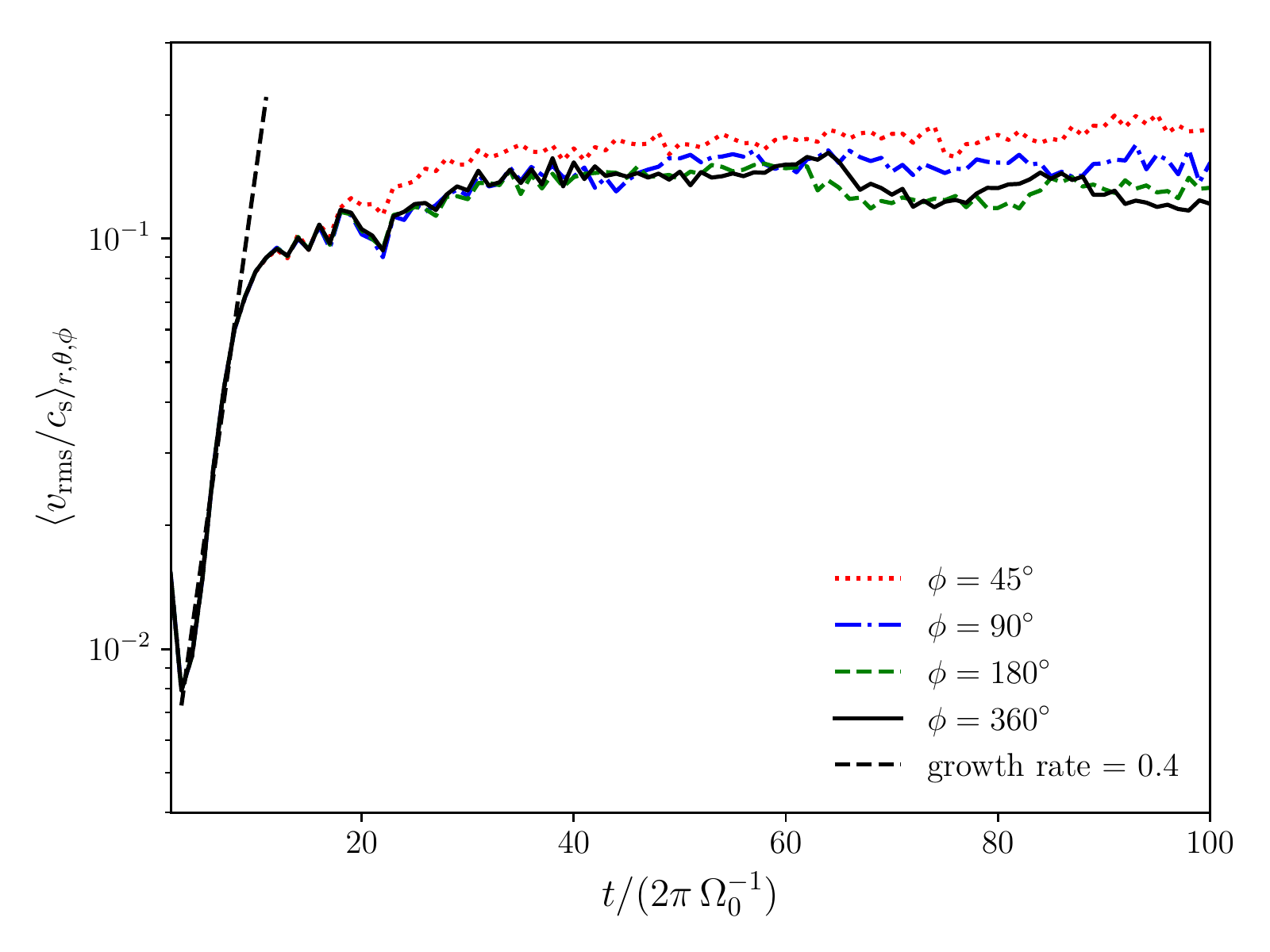}
\caption{Time evolution of the RMS-velocity of the gas. The values are averaged over the radial and azimuthal subdomain. We observe rapid turbulent growth at the onset of the simulation and saturation after a few 10 orbits. We fit the exponential growth rate of the velocity as 0.4 per orbit.}
\label{fig:vrmsvT}
\end{figure}

\revisedvtwo{We consider only the radial and vertical component of velocity because the disk rotation profile is not keplerian in this work and spatially varying systematic deviations cannot be taken into account.} The results are presented in figure \revised{\ref{fig:vrmsvT}}. We find the values of all runs to agree with each other for the first few orbits as expected, as the limit in azimuthal wavenumber does not influence the onset of the instability. The overall onset of the instability is observed earlier than in other studies. This is explained by the specific choice of parameters in our setup, which allows for earlier onset of the instability due to the very short cooling time. The saturated values for cases missing the lower azimuthal wave-numbers p45 and p90 are higher than for the other cases. 
We measure the growth rate of the velocity perturbations as 0.4 per orbit, about double the value reported by \cite{Stoll+Kley2014}, which is in \revised{excellent} agreement with
theory as growth rate is proportional to the pressure scale height $\sigma \sim q \Omega \frac{H}{R}$ \citep{Nelson+2013}, which was only half the value we adopted. 
Note that \cite{Stoll+Kley2014} give their value for the growth rates for the kinetic energy. 
\revised{Comparing the time evolution of the rms-velocity to the time evolution of the alpha value in figure \ref{fig:alphavTime}, we find that the rms-velocity saturates after about 20 orbital periods, while $\alpha$, the quantity related to angular momentum transport, saturates only after 100 orbital period. This behaviour is linked to the transport of energy to larger scales in the disk and we will discuss this further in section \ref{sec:discRWI}.}

\begin{figure}
\centering
\includegraphics[width=\columnwidth]{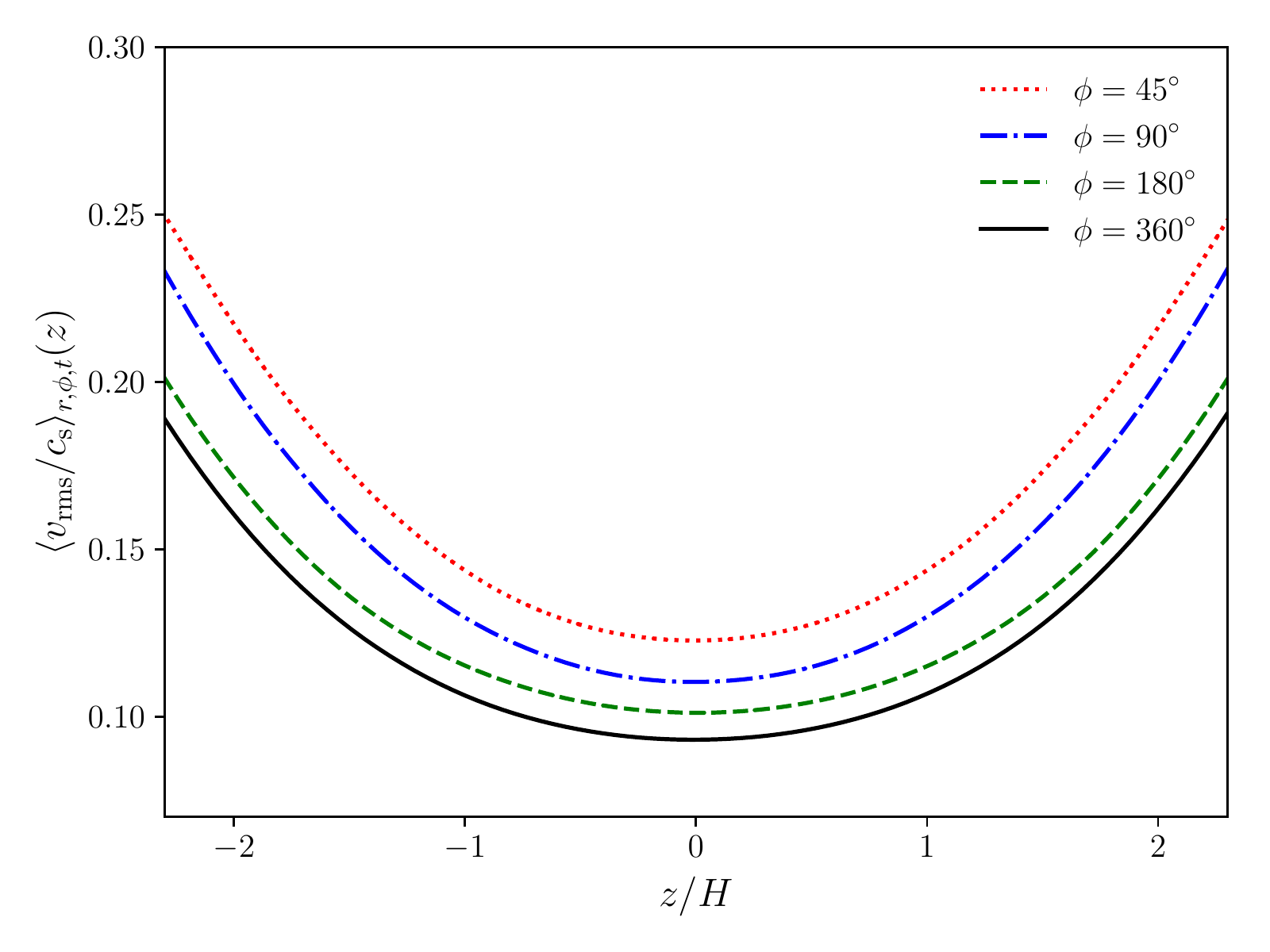}
\caption{Meridional profile of the vrms value. The vrms values are averaged over the radial and azimuthal subdomain and between 500 and 700 orbits. We again observe significantly lower values for the run p360 than for the other runs.}
\label{fig:vrmsvR}
\end{figure}
Figure \ref{fig:vrmsvR} shows the $v_\mathrm{rms}$-values as a function of height averaged over the radial and azimuthal subdomain and between 500 and 700 orbits. We find a similar shape for the vertical structure of all runs with low velocities in the midplane and rising with z/H. We find a systematic positive offset for all runs compared to p360 with p45 having the largest offset. This compares to our findings for the $\alpha$ viscosity parameter in figure \ref{fig:alphavR}, suggesting a smaller $\phi_\mathrm{max}$ systematically overestimates the turbulence strength. 

\begin{figure}
\centering
\includegraphics[width=\columnwidth]{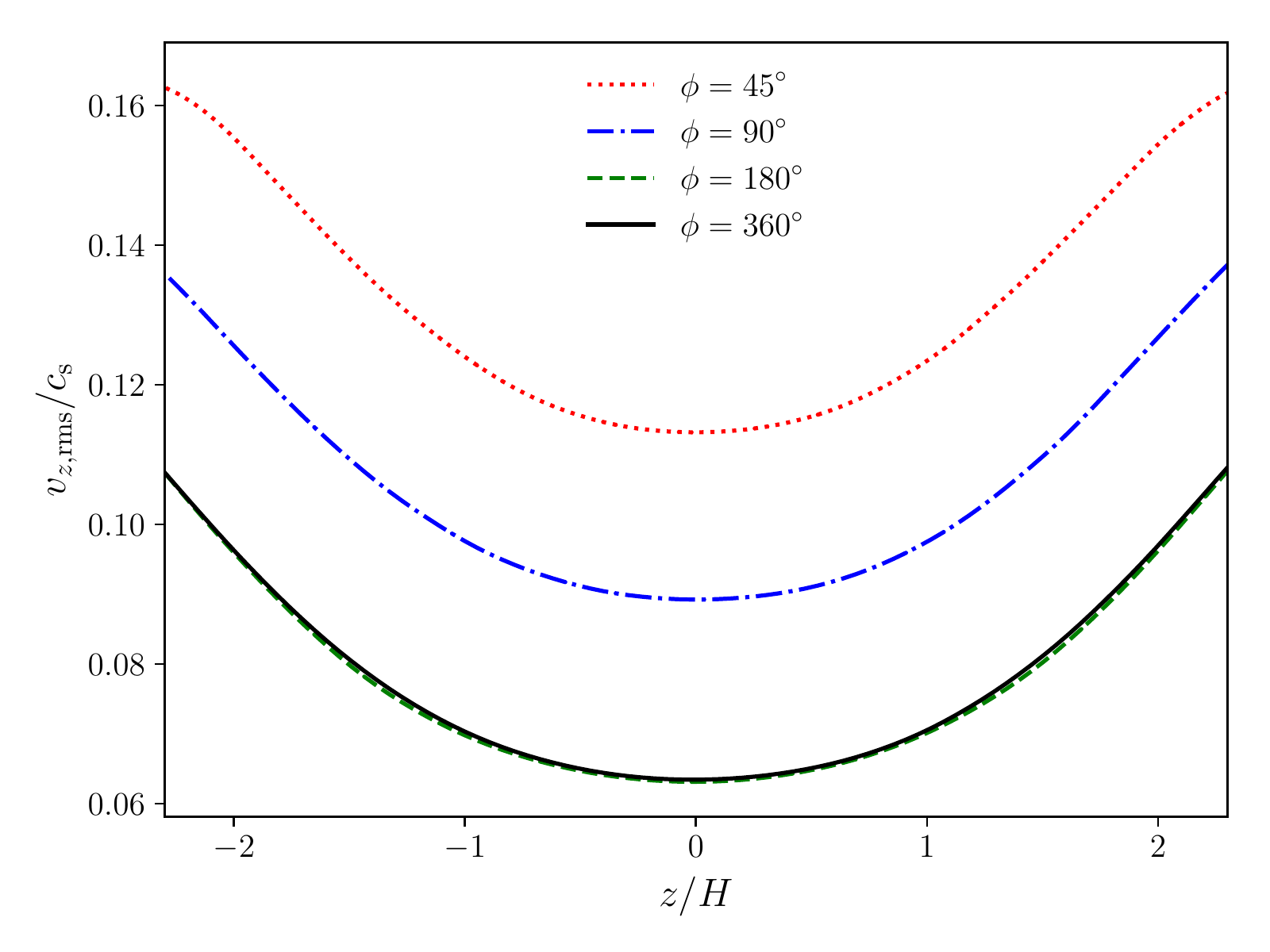}
\caption{Vertical profile of the vertical rms velocity. We average the values over the radial and azimuthal subdomain and between 500 and 700 orbits. We find a large offset for the cases p45 and p90 compared to p180 and p360, which are in good agreement with each other.}
\label{fig:vZrmsvz}
\end{figure}

In spectral line observations of protoplanetary disks the total rms velocity can however not be measured. The most easily accessible quantity is the vertical component of the velocity, which can be measured by determining the line broadening of face-on disks. We separately calculate the height profile for this quantity in figure \ref{fig:vZrmsvz}. We again see the large offset for the simulations p45 and p90 compared to the case p360, the case p180 shows similar values to the $360^\circ$ case. The height profile for all simulations follows a similar shape as the total $v_\mathrm{rms}$ values with low values in the midplane growing with height z.

\begin{figure}
\centering
\includegraphics[width=\columnwidth]{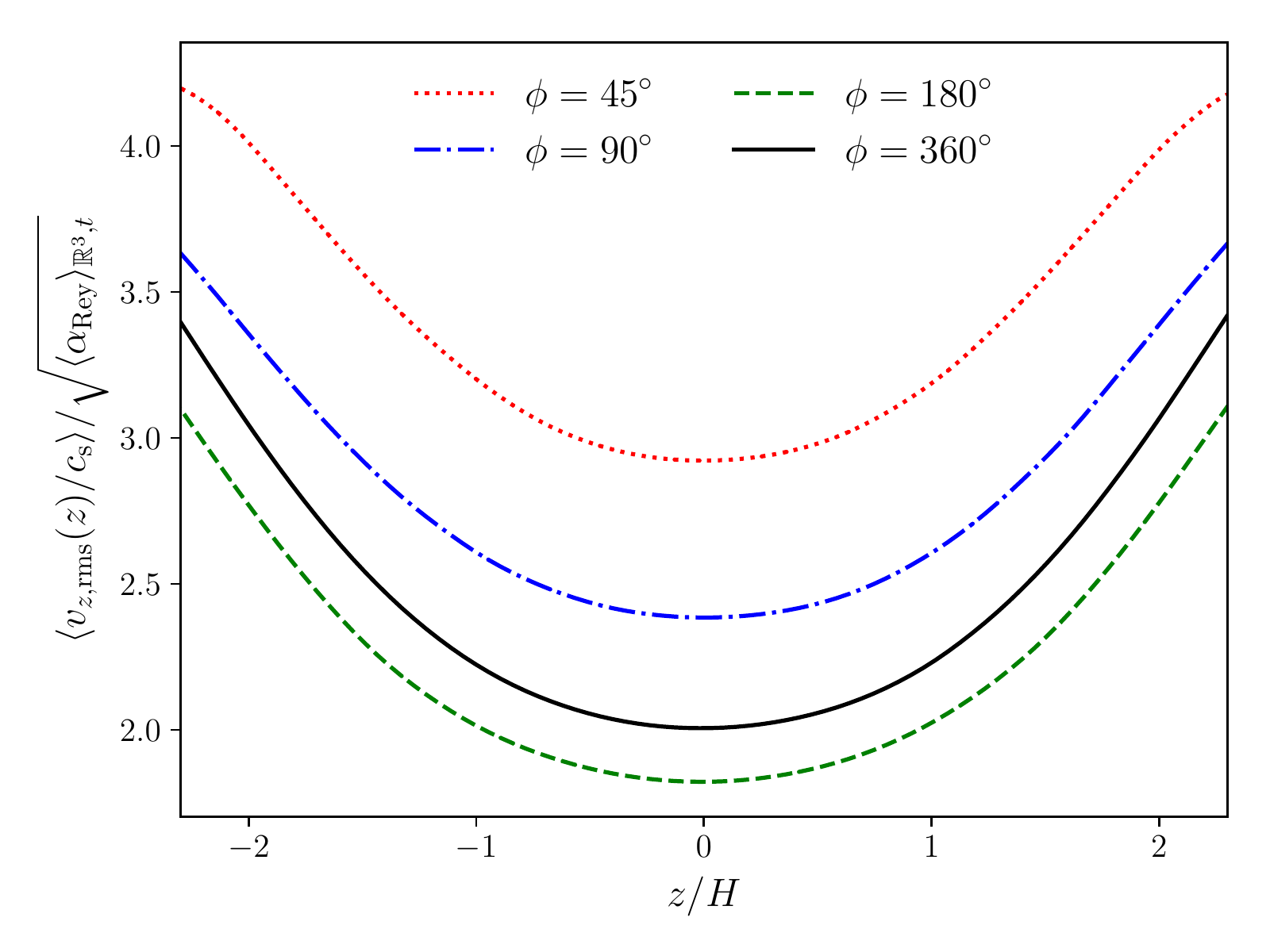}
\caption{Vertical profile of the vertical rms velocity  divided by the square root of the total alpha value of the disk (see table \ref{tab:runs}). We average the values over the radial and azimuthal subdomain and between 500 and 700 orbits. All simulations show an offset from unity, so estimating the $\alpha$-values from $v_{z,\mathrm{rms}}$ leads to an overestimation. We again find a large positive offset for the cases p45 and p90 compared to p360, whereas p180 has a smaller negative offset with respect to p360.}
\label{fig:vzrmsVsAlpha}
\end{figure}

\cite{Cuzzi+2001} related the rms-velocity to the turbulent viscosity parameter $\alpha$ via
\begin{equation}
v_\mathrm{rms} = \sqrt{\alpha} c_\mathrm{s}
\label{eqn:vrmsValpha}
\end{equation}
(their equation 2) if the largest eddies have a rotation frequency comparable to the orbital frequency. To check the applicability of this relation to the turbulence induced by the VSI, we plot the ratio of the height dependent z component of the rms-velocity to the square-root of the simulation averaged alpha value (see table \ref{tab:runs}) in figure \ref{fig:vzrmsVsAlpha}. 
For all simulations, the ratio of $v_\mathrm{rms}/c_\mathrm{s}$ to $\sqrt{\alpha}$ is above unity. Therefore the angular momentum transport in our simulations of the VSI is \revised{weaker than one would expect from the measured velocities}. The deviation depends on height with values closer to unity in the midplane. The overall deviation is largest again for the case p45, which also showed the highest values for $v_\mathrm{rms}/c_\mathrm{s}$ and $\alpha$ and decreases with $\phi_\mathrm{max}$. For the case of a full disk we find however values larger than for $\phi_\mathrm{max}=180^\circ$. For protoplanetary disks subject to the VSI, the $\alpha$ values calculated from measured turbulent velocities should therefore be treated with caution. Depending on the height above the midplane where the measurement is taken, the values for $\alpha$ could be overestimated by up to one order of magnitude. Also, when comparing the alpha values of simulations and observations, the influence of the azimuthal extent of the simulation should be taken into account.

\subsection{The influence of $\phi_{max}$ on the disk structure}
\begin{figure*}
\centering
\includegraphics[width=\textwidth]{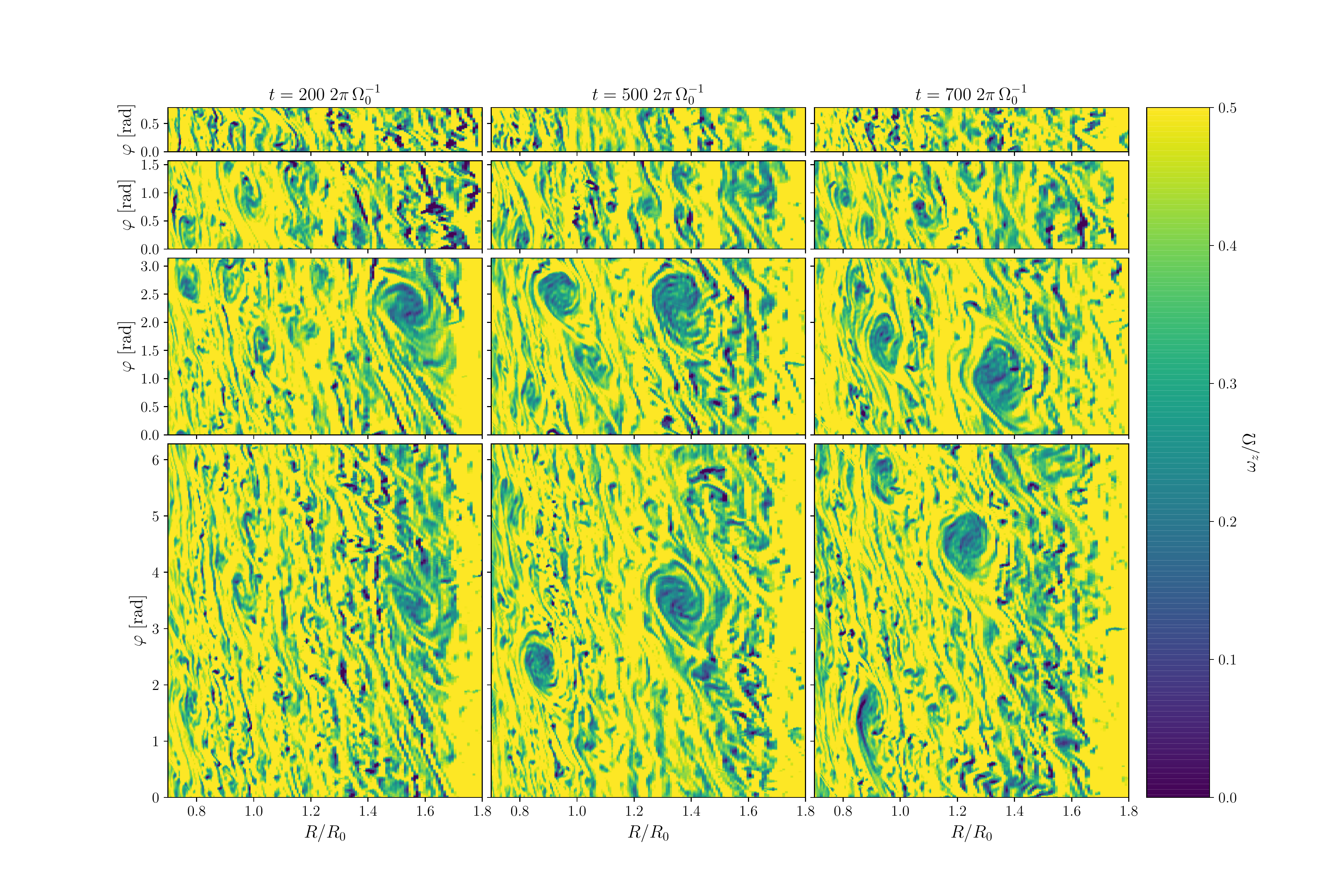}
\caption{The midplane value of the z-component of vorticity after 200, 500 and 700 local orbits into the simulation. The azimuthal size of the simulation increases from top to bottom. For the upper two rows we see the formation of zonal flows and small vortex structures. In contrast to this we see large vortices in the lower rows for the second and third snapshot.}
\label{fig:vorticity}
\end{figure*}
To asses the influence of $\phi_\mathrm{max}$ on the outcome of the numerical simulation, we plot the z component of the vorticity
\begin{equation}
\omega_z = \left(\bm{\nabla} \times \bm{v} \right)_z
\end{equation}
in the disk midplane for three different times: after 200, 500 and 700 local orbits. 
The results are presented in figure \ref{fig:vorticity}. For the case of p45, we find small vortices with aspect ratios $\chi \approx 4$ and zonal flows. All structures show strong variations in time, as can be seen by the differences in the time frames shown in figure \ref{fig:vorticity}. These structures also appear in the case of p90, where also larger structures similar to vortices emerge in the first frame but are destroyed again in the second frame of figure \ref{fig:vorticity}, also pointing to high variability with time.
This changes for the simulations p180 and p360. Here we observe small, unstable vortices forming quickly in the beginning and additionally two larger vortices with aspect ratios $\chi \approx 8 - 10$ after a few hundred orbits, seen in the left frame in figure \ref{fig:vorticity}. The large vortices continue to appear both after 500 and 700 orbits (fig. \ref{fig:vorticity}, middle and right columns), suggesting stability over larger times. The middle and right column also show additional large vortices appearing at later times.
This change in overall structure for azimuthal extents larger than $180^\circ$ suggests that the approximation of large m restricts the development of large, long lived structures in VSI disks. 

\begin{figure*}
\centering
\includegraphics[width=\textwidth]{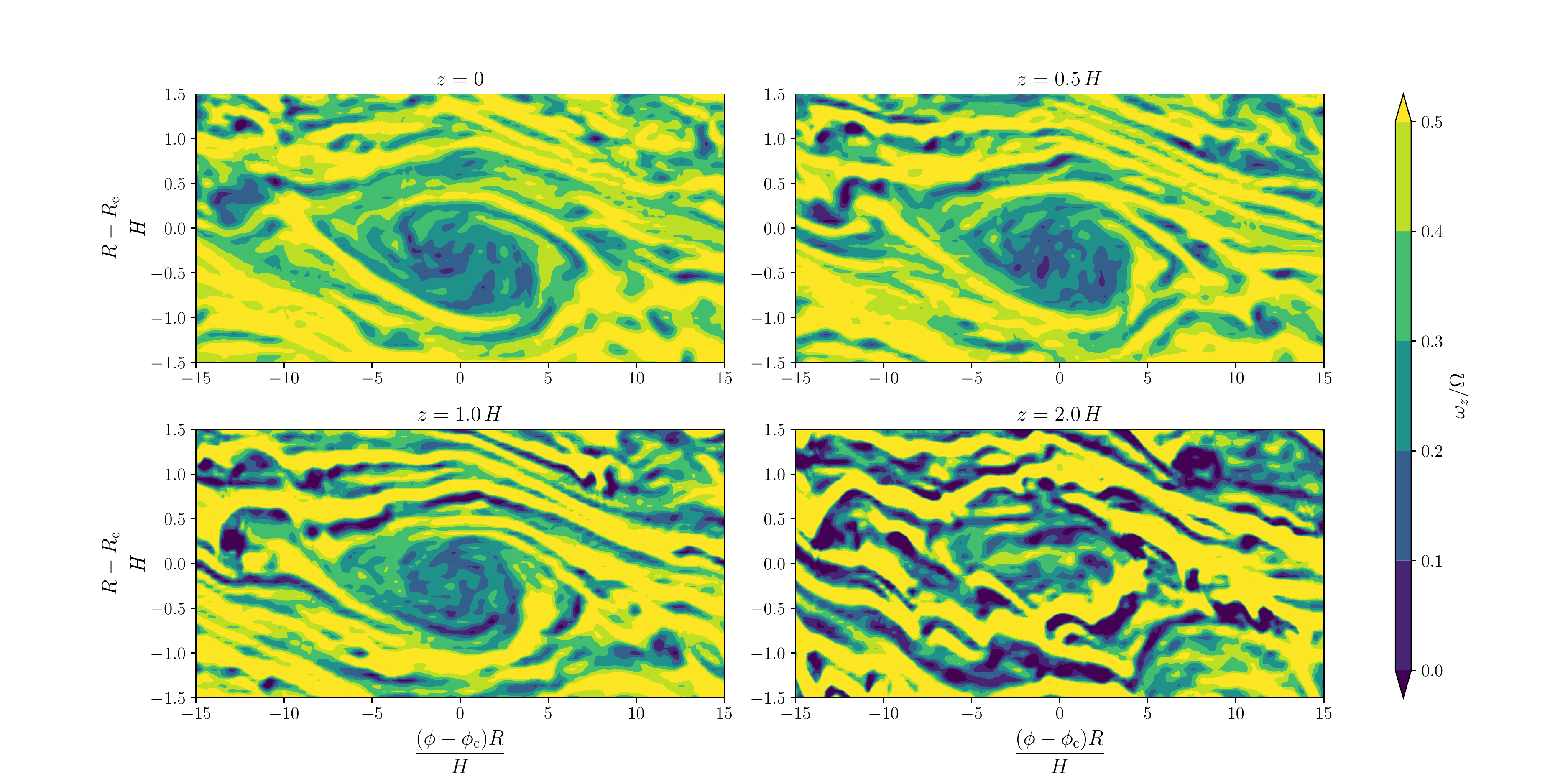}
\caption{Contour plots of the vorticity at different heights from the midplane for the simulation p360 at the position of the outer vortex $(R_\mathrm{c},\phi_\mathrm{c}) = (1.4\,R_0,3.5)$ after 500 orbits. The plots show the highly turbulent state in the vortex.}
\label{fig:vortZoom}
\end{figure*}

\begin{figure*}
\centering
\includegraphics[width=\textwidth]{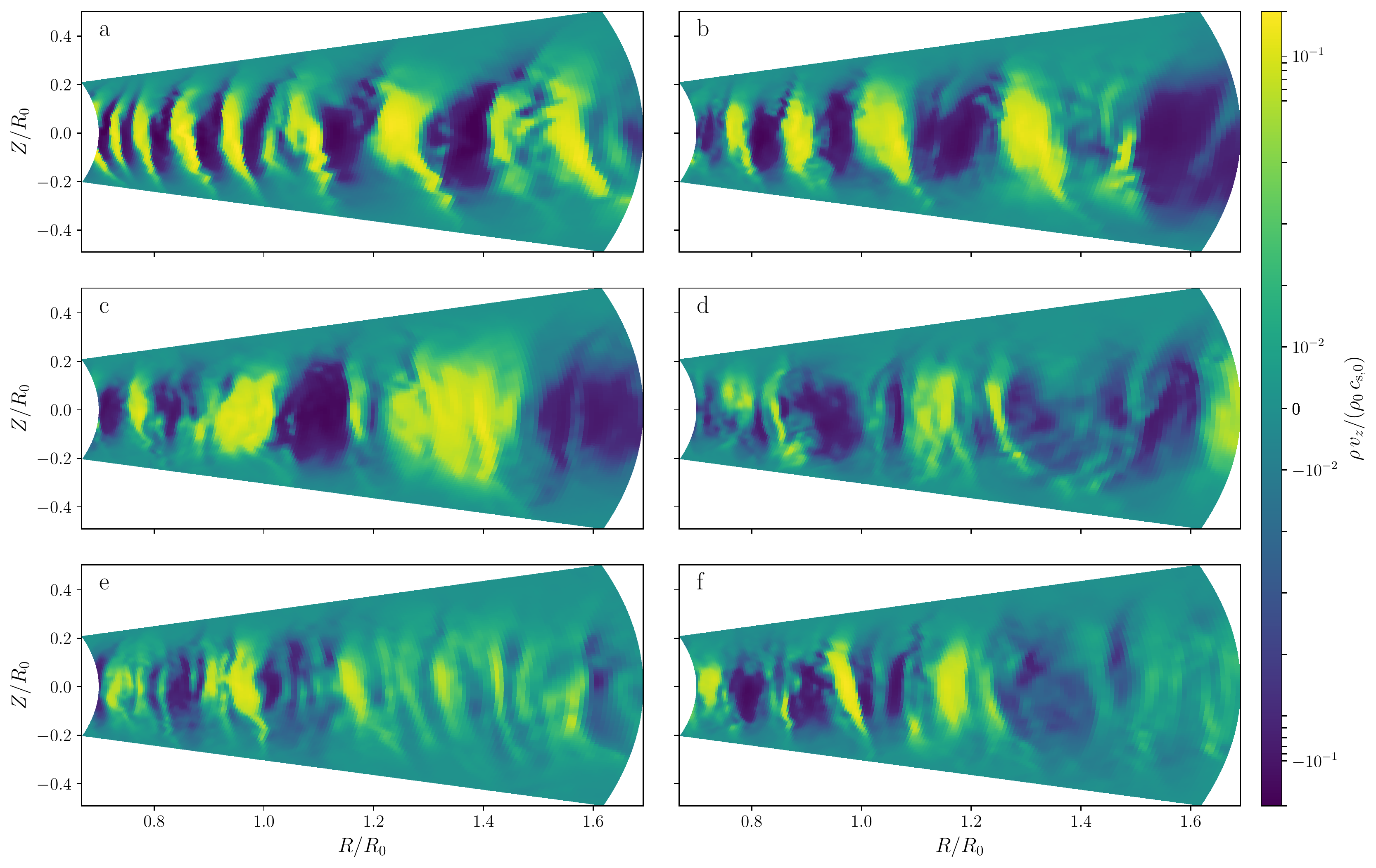}
\caption{Vertical slice of the vertical velocity in the disk for different azimuthal ranges after 500 orbital periods. Plot (a) shows a 2D axisymmetric disk after 80 orbits for comparison, (b)-(d) the models p45, p90 and p180 respectively and (e) and (f) the model p360, where (e) is taken at a $\varphi$ position without a vortex and (f) at the center of the outer vortex shown in figure \ref{fig:vorticity} (middle column). The color scale uses linear scaling for absolute values smaller 0.05 and logarithmic otherwise to enhance visibility. }
\label{fig:vzRz}
\end{figure*}

Looking at the large vortices specifically, we find them to have an inner turbulent structure, similar to the one found in \citep{Raettig+2013,Lyra2014}. Figure \ref{fig:vortZoom} illustrates this, showing the structures inside the outer vortex of the simulation p360 at $(R_\mathrm{c},\phi_\mathrm{c}) = (1.4\,R_0,3.5)$ after 500 orbits (fig. \ref{fig:vorticity}, middle column). We again plot the z component of vorticity, now for 4 different heights z = 0, 0.5 H, 1 H and 2 H above the midplane of the disk. We multiply the azimuth with the central radius and scale both axis with the local pressure scale height to show the true size of the vortex. 

The structure inside the vortex shows changes on small scales, forming a highly turbulent substructure. The turbulent structure is visible throughout the upper and the lower left panel of figure \ref{fig:vortZoom}. In these 3 panels we also find a consistent outer boundary for the vortex as an azimuthally elongated radially narrow region of higher vorticity relative to the inside of the vortex, forming a shell around the vortex. This structure is consistent with the predictions of \cite{Lesur+Papaloizou2009} for the elliptic instability, which we will discuss in section \ref{sec:disEI}. The lower right panel of figure \ref{fig:vortZoom} shows the vortex structure fading into the turbulent background structure of the disk, suggesting the vortex extends to about 2 scale heights above and below the midplane.

\begin{figure*}
\centering
\includegraphics[width=\textwidth]{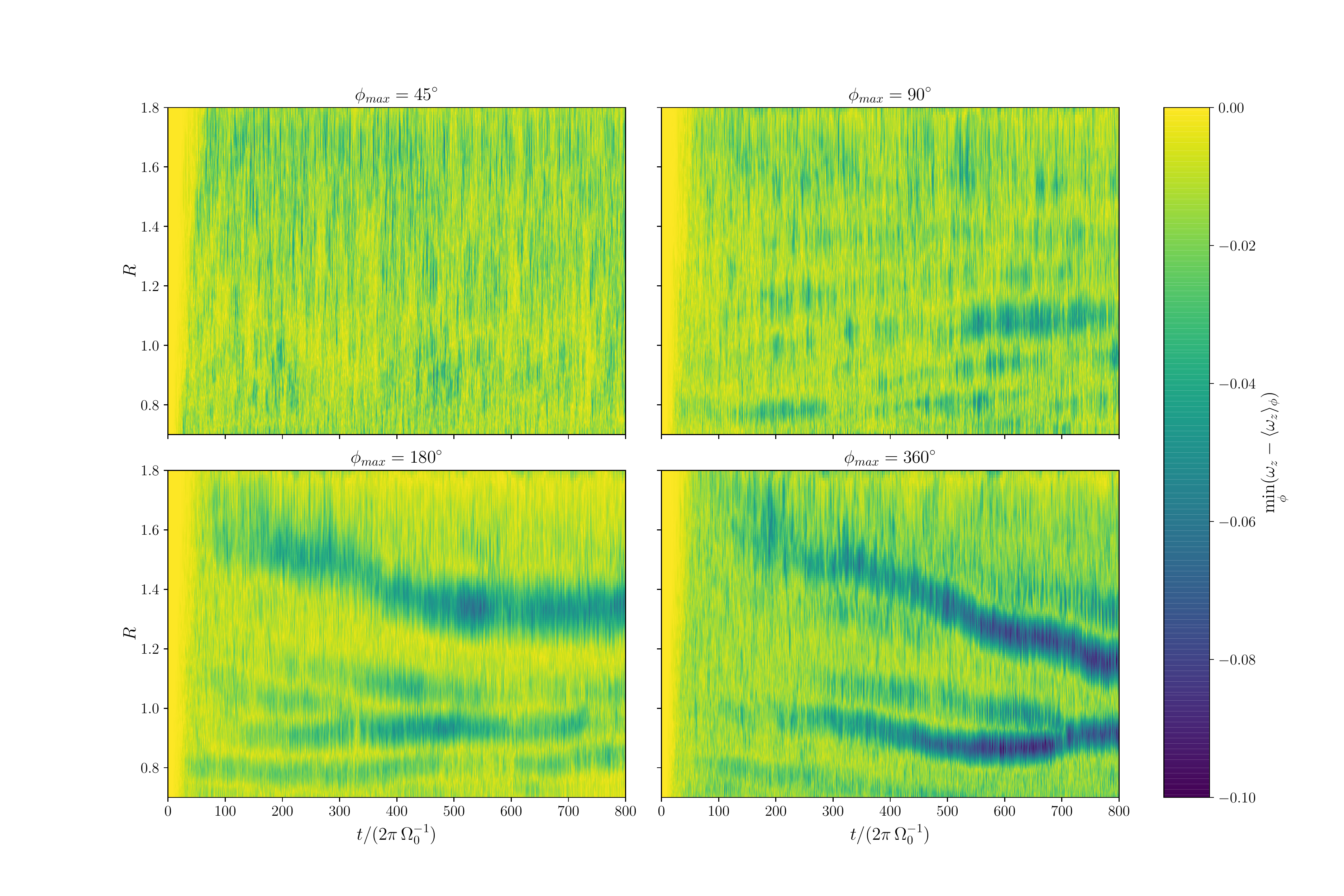}
\caption{Track of the vortex radial position over time via a function of $\omega_z$. Only the cases p180 and p360 in the lower panels show vortices surviving over hundreds of orbits, indicated by the darker lines. The lower panels also show the position of the vortex moving to smaller radii with time.}
\label{fig:vortRvT}
\end{figure*}

\revised{
To asses the influence of the forming vortices on the flow structure reported for the VSI we plot the vertical velocity times density in the R-Z plane in figure \ref{fig:vzRz} after 500 orbits, corresponding to the middle column of figure \ref{fig:vorticity}. Figure \ref{fig:vzRz} (b) and (c) show the models p45 and p90 at $\varphi = \pi/8$ and $\varphi = \pi/4$ respectively and  (d) shows the model p180 at $\varphi = 2.5$, the center position of both vortices at that time. Figure \ref{fig:vzRz} (e) and (f) show model p360 at 2 different $\varphi$ positions: (e) at $\varphi = 1.0$, where no large vortices are intersected, and (f) at $\varphi = 3.6$, the center position of the outer vortex. Figure \ref{fig:vzRz} (a) shows the 2D axisymmetric simulation with the same r,z resolution from the resolution study shown in appendix A after 80 orbits. The snapshot from the 2D run is taken at an earlier time to avoid the non-linear phase of the axisymmetric VSI, which differs from the full 3D simulations due to the occurrence of the RWI in 3D.
}

\revised{
We find a similar vertical velocity profile as reported by \cite{Nelson+2013},\cite{Stoll+Kley2014} and \cite{Flock+2017} with strong vertical motions present over the whole height of the disk. We find the run p45 agreeing well with the early stage of the 2D comparison simulation, showing a pattern of disk annuli of $\Delta R \approx 0.5 H $ with alternating positive and negative vertical velocity and symmetry about the midplane of the disk. In (c) we see the overall pattern still preserved for the case p90, but the annuli now have a larger radial width of $\Delta R \approx H $. This picture changes once the disk is able to form vortices. Fig. (d) shows the case of p180 at the centre position of the inner vortex. The ordered pattern seen in (a)-(c) is broken at the radii of the vortices, although it is still partly visible at other radii. The size of the annuli of alternating velocity also decreases in this simulation. Both vortices have negative vertical velocities, although the velocity is greatly reduced with respect to the surrounding, especially for the outer vortex. Figure (f) shows a similar behaviour for the case p360. The overall structure of the disk is less ordered than for the axisymmetric case and the vortex has a lower vertical velocity as the surrounding disk, although a similarity with the VSI velocity pattern is retained. This is also true for slices at $\varphi$ positions outside the vortex, shown in (e). There we also find the pattern of positive and negative $v_z$ annuli, although their strength and width again is diminished with respect to the axisymmetric case. Interestingly, we also find a turbulent ring structure at the raidal position of the outer vortex in (e). This could indicate the vortex is embedded in a larger flow structure or the occurrence of small vortices which are elongated in z-direction. Comparing (e) and (f) in the inner part of the disk, we find the radial positions of the annuli described above do not coincide, indicating the annuli are not axisymmetric anymore for these larger disks.
}

\subsection{Vortex lifetime}
For planetesimal formation the lifetime of a vortex is a critical factor determining the trapping efficiency. Therefore we are interested in the lifetime of the vortices generated by the instability. A distinctive signature of a large anticylonic vortex is a minimum in local vorticity stretching out over a fraction of a disk annulus.
We therefore apply a box filter in radial and azimuthal direction to the midplane z-vorticity values. In azimuth, we apply a filter width of \revised{48 and 96 grid cells for p45 and p90 respectively and 200 grid cells for both p180 and p360}. In radial direction, the filter width is 30 grid cells,\revised{equalling $2H$}. \revised{The filter damps fluctuations on scales smaller than the filter width, enhancing the visibility of large scale structures.} To outline the radial position of the large vortices, we then find the minimum in the deviation of the smoothed $\omega_z$ from the azimuthal average in the respective annulus. The calculation of the deviation from the azimuth ensures we exclude the large azimuthal flow structure occurring at the outer boundary in figure \ref{fig:vorticity}. Figure \ref{fig:vortRvT} shows the evolution of this value as a function of radius and time, outlining the radial position of the vortices in the disk over the run of the simulation. 

We observe multiple long living vortices emerging in run p180 and two in run p360, seen in the lower panel in figure \ref{fig:vortRvT}. They start with distances of 1 to 5 pressure scale heights and are observed to migrate inwards with a migration rate lower than 0.0005 $R_0/\Omega$. \revised{\cite{Paardekooper+2010} find similar migration rates for vortices in laminar 2D vertically integrated disks. Though their results are not directly transferable to the case presented here, we think the underlying mechanism is the most plausible explanation of the phenomenon observed.}
The outward migration of the inner vortex in p360 can be explained by the occurrence of a surface density maximum moving outward. The vortex is trapped in the maximum and is dragged outward again, as it cannot drift across it \citep{Paardekooper+2010}. Figure \ref{fig:vortRvT} in conjunction with fig.\ref{fig:vorticity} also shows vortices interacting and after some time merging once one vortex migrates within one pressure scale height radial distance to another.

In contrast, we don't observe the same in the simulations p45 and p90, shown in the upper row of figure \ref{fig:vortRvT}. 
We do not see any sign of larger vortices in simulation p45, which is in agreement with \citep{Richard+2016}. The case p90 shows a vortex forming after around 150 and another after 550 and existing at least intermittently. This suggests the phi range of a simulation to be crucial in the development and sustaining of large, long lived vortices.

\section{Discussion} 
\label{sec:discussion}

\subsection{RWI as secondary Instability}
\label{sec:discRWI}
In this section, we now focus on evidence in our simulations that suggests the triggering of the Rossby-Wave-Instability (RWI) to initiate vortices. We therefore calculate the values of the critical function $\mathscr{L}$ as given in equation \ref{eqn:RWI} from an azimuthally averaged density, pressure and velocity structure of our simulation. We also averaged the field quantities over $\theta$ before calculating $\mathscr{L}$.

\begin{figure*}
\centering
\includegraphics[width=\textwidth]{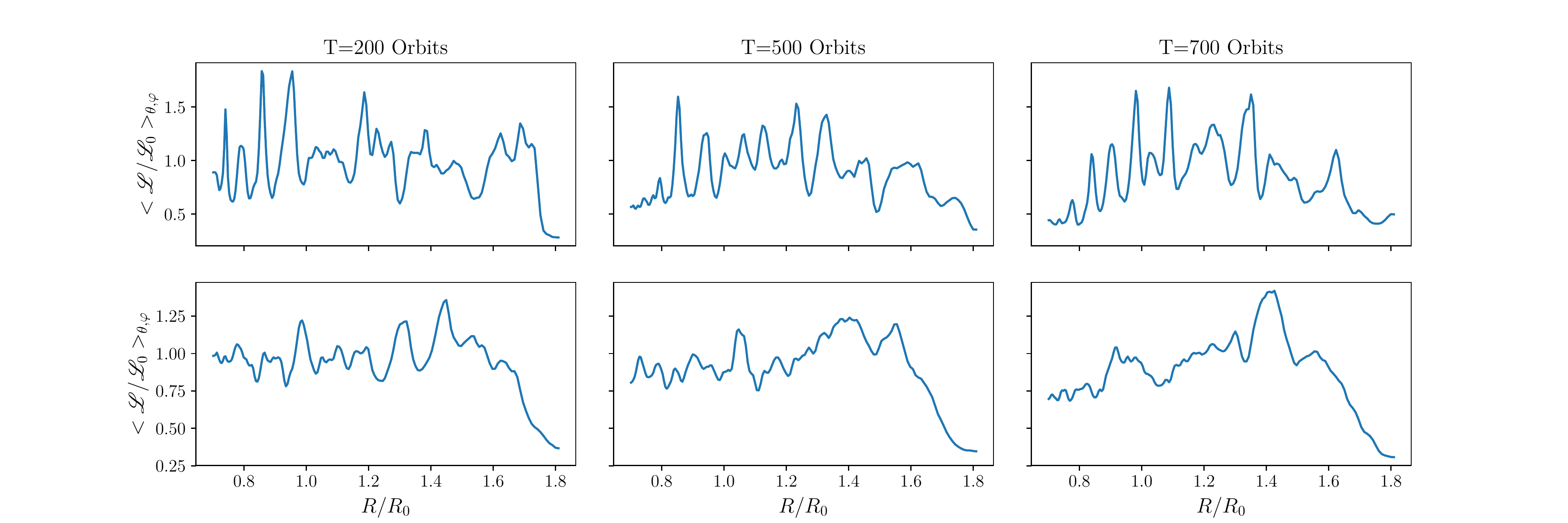}
\caption{The critical function $\mathscr{L}$ of the RWI for 3 different times normalized by the initial profile. The upper row represents the run p45 and the lower represents run p360.}
\label{fig:RWIL}
\end{figure*}

Figure \ref{fig:RWIL} shows $\mathscr{L}/\mathscr{L}_0$ as a function of radius after 200, 500 and 700 orbits. In the top row we show the results from p45. All plots show multiple strong global maxima, suggesting the disk to be in principle susceptible to the RWI. The amplitude of the extrema decreases with time but they stay as distinctive features throughout the simulation time. In the bottom row we show  $\mathscr{L}/\mathscr{L}_0$ for p360. Here we also find extrema, but fewer in number and not as distinct as for p45. Also, the overall amplitude of the extrema is lower as in the other case.

Taking a closer look at the lower right plot in figure \ref{fig:RWIL}, we find the most distinctive peaks at R=1.0,1.3 and 1.45. Comparing these to the initial locations of the vortices we find in the lower left subplot of figure \ref{fig:vortRvT}, we find them to be in good agreement. This is clear evidence that the RWI is triggering the large vortices we observe in our simulations. The decrease in amplitude of the extrema after saturation of the instability has been shown for artificial RWI vortices by \citep{Meheut+2010} and can explain the decrease we witness in the p360 case.


\begin{figure}
\centering
\includegraphics[width=\columnwidth]{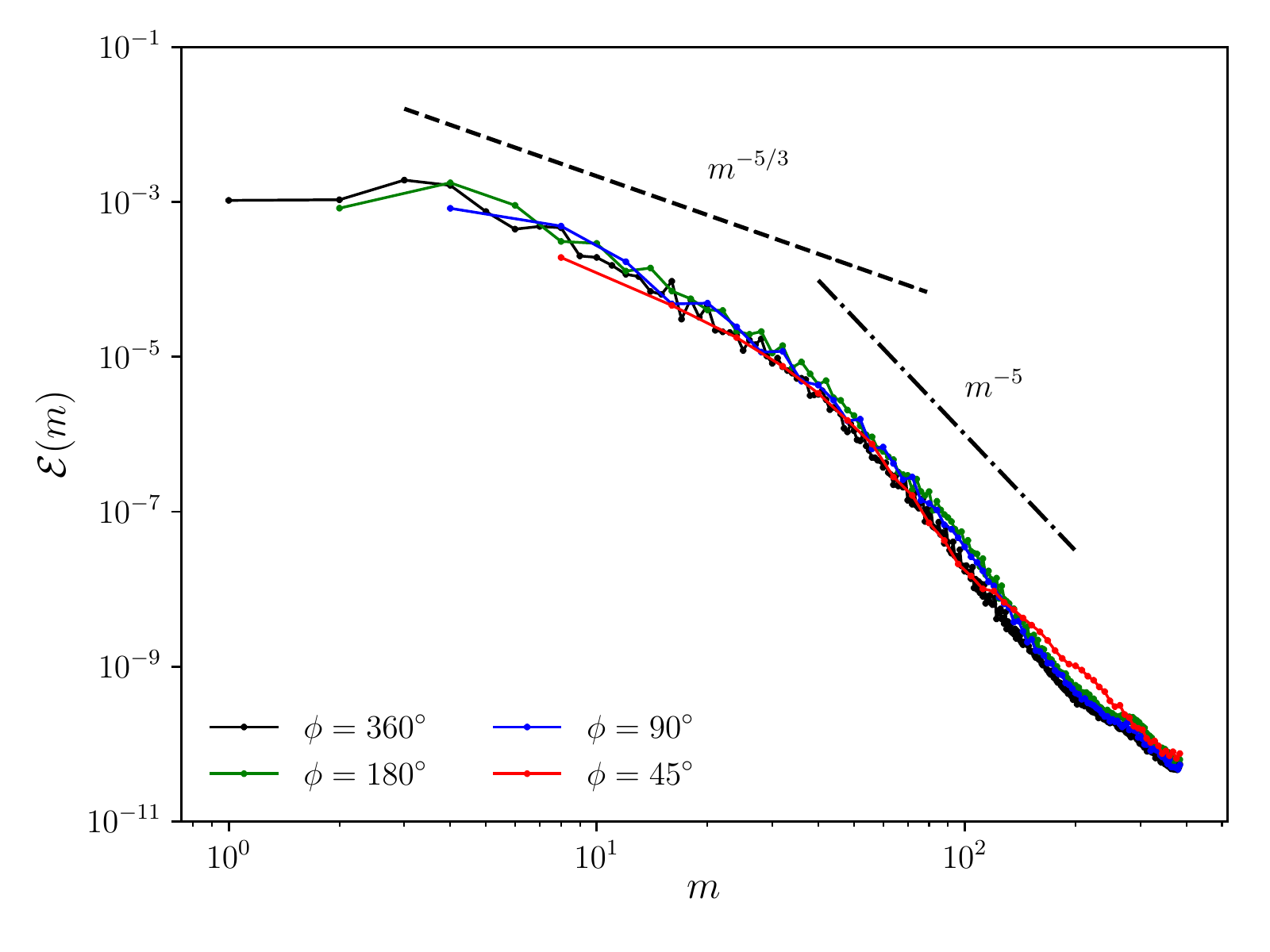}
\caption{Kinetic energy spectrum at 700 orbits. Note the distinctive broken power law behaviour characteristic for 2D turbulent fluid flow with \revised{both inverse energy and downward enstrophy cascade}. For comparison we plot the power laws $\mathcal{E}(m)\propto m^{-5/3}$ and $\mathcal{E}(m)\propto m^{-5}$, indicators of the inverse energy and \revised{downward} enstropy cascade respectively. }
\label{fig:RWIEnergy}
\end{figure}
To investigate this, we also look at the azimuth of the kinetic energy spectrum in the midplane of the disk.
\begin{equation}
\mathcal{E}(m) = 0.5 \sum\limits_i \langle\vert(\mathcal{F}(v_{i,z=0}))^2\vert\rangle_r \qquad i \in [r,\theta,\phi]
\end{equation}
$\mathcal{F}$ denotes the Fourier transform of the respective velocity component.

\citep{Li+2000} showed the RWI has a maximum growth rate for the azimuthal wave number m in the range of $m = 3-6$. These wavenumbers can only be accessed for simulations with an azimuthal extent equal or larger than 90 degree. Therefore the RWI is expected to only grow inefficiently in the p45 simulation. Figure \ref{fig:RWIEnergy} shows the radial average of the azimuthal component of the kinetic energy spectrum after 700 orbits.

We find a broken power law for $\mathcal{E}(m)$ in all our simulations. At smaller wavenumbers m, the slope falls as $m^{-\frac{5}{3}}$, as expected for the upward Kolmogorov cascade in a rotationally dominated
flow above the Rhines scale \citep{Rhines1975}, i.e. where the eddy turn over time is longer than the rotational period of the system.
At large wavenumbers, the spectrum behaves $\propto m^{-5}$, a scaling in agreement with a 2D enstrophy downward cascade. This result indicates that our simulations are predominantly 2D/rotation dominated due to the fast rotating flow in the disk.\revised{This is supported by our finding that the Reynolds stresses saturate at a later time than the rms-velocity. The rms-velocity directly traces the turbulence generated at smaller scales induced by the VSI, but $\alpha_\mathrm{Rey}$ traces the angular momentum transported at larger scales. The difference in saturation time is then explained with the observed inverse energy cascade in the disk, which has to transport the energy generated at small scales to the larger scales on which angular momentum is transported in the disk.} \revised{This is supported by the fact that we find from figure \ref{fig:RWIEnergy}} that the energy injection of the VSI occurs at about $m=40-60$, which is about $1-1.5$ pressure scale heights. 

At the \revised{smallest} wavenumbers $m$, we find a maximum for $\mathcal{E}(m)$ at $m=3$ and $m=4$ for the cases with $\phi_\mathrm{max}=180^\circ$ and $360^\circ$, whereas for the other cases the energy piles up at $m=4$ and $m=8$ for $\phi_\mathrm{max}=90^\circ$ and $45^\circ$, the largest respective wavenumber accessible. For the cases with $\phi_\mathrm{max}\leq 180^\circ$, the large vortices formed by the RWI extract energy from the flow, whereas the absence of large vortices in the other simulations forces the energy to be deposited in the largest mode accessible in the simulation. This can also explain the systematically higher rms-velocities we found for $\phi_\mathrm{max}=45^\circ$ and $90^\circ$ in figures \ref{fig:vrmsvT} and \ref{fig:vrmsvR}.

To confirm we are in the rotation dominated regime in our simulations, we plot the Rossby number
\begin{equation}
\mathrm{Ro} = \frac{u}{l\cdot \Omega_0}
\label{eqn:RossbyNr}
\end{equation}
with $u$ being the flow velocity at length scale $l$. The Rossby number gives the relative importance of Coriolis forces vs. inertial forces and is smaller unity if Coriolis forces are non-negligible for the flow in the system. We can express this number also as a function of wavenumber $m$ using the kinetic energy spectrum to define the velocity spectrum.

\begin{equation}
\mathrm{Ro}(m) = \frac{\sqrt{m/(2\pi \,R_0)\mathcal{E}(m)}}{m/(2\pi \,R_0)\cdot \Omega_0}
\end{equation}

\begin{figure}
\centering
\includegraphics[width=\columnwidth]{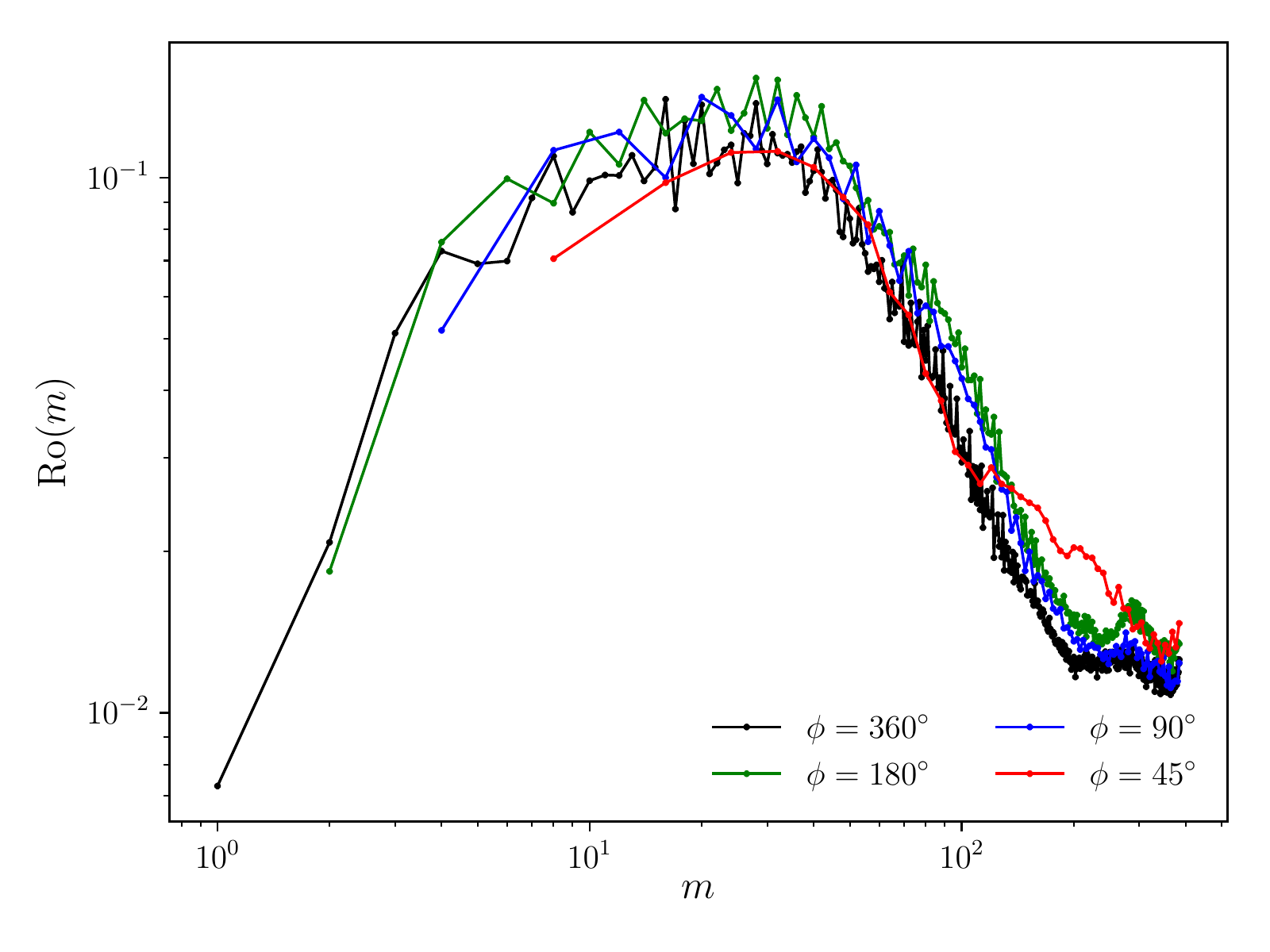}
\caption{Rossby number as a function of azimuthal wavelength at 700 orbits. We find Ro to rise at large scale and later fall towards larger scales, with a maximum at the energy injection scale. The Rossby number stays below unity at all scales resolved in our simulation.}
\label{fig:RoVsM}
\end{figure}
We plot $\mathrm{Ro}(m)$ as a function of azimuthal wavenumber $m$ in figure \ref{fig:RoVsM}. We find the Rossby number stays below unity at all scales accessible in our simulations, confirming we are indeed rotationally dominated. We also find Ro $\approx$ const. at the largest wave numbers $m$, indicating a change of slope for $\mathcal{E}$ not prominently visible in figure \ref{fig:RWIEnergy}. 
Of course we only investigate one component of the turbulent spectrum, because neither the radial nor the vertical wave numbers are as easy measurable as the azimuthal component. 
VSI is known to favour high radial and low vertical wave-numbers, yet this investigation was not feasible with the current data sets, but shall be attempted in the future. In any case
the low values for Ro are an indication that all our turbulence is strongly rotational dominated and thus most likely not isotropic anyway. 

Because Ro $< 1$ in our simulation, we now revisit equation \ref{eqn:vrmsValpha} . \cite{Cuzzi+2001} developed the equations under the assumption that the disk turbulence is isotropic and follows a Kolmogorov dissipation law, which implies that turbulence has to be in the Ro $>1$ regime, which is clearly not the case in our work. From equation \ref{eqn:RossbyNr} we can estimate the wavenumber at which our system should satisfy this condition. For a typical velocity $v_\mathrm{rms} = 0.01 v_\mathrm{k,0}$ we get
\begin{equation}
m_{\mathrm{Ro} = 1} = \frac{v_\mathrm{rms} \cdot m}{2\pi\, R_0} \approx 600 \frac{1}{R_0}  \: ,
\end{equation}
a value much larger than the Nyquist-wavenumber of our setup. This suggests that the rms-velocities in the disk act on much larger scales than the turbulence scales, and the ansatz presented in \cite{Cuzzi+2001} is not always valid in protoplanetary disks. We however consider only the azimuthal direction in our analysis, leaving the possibility of a different scaling in the other spatial directions.\revised{ Nevertheless, even if there might be more energy at large radial and vertical wavenumbers $k_R,k_z \equiv 600/R$, this still means that turbulence is also on small scales highly anisotropic, asking for revisiting the Kolomogorov picture for small scale turbulence in protoplanetary disks.} Also, the results of section \ref{sec:results} show the ansatz works quite well despite $v_\mathrm{rms}$ being generated on different scales than assumed for the turbulent $\alpha$.


\subsection{Influence of elliptic instability}
\label{sec:disEI}
In section \ref{sec:results} we described a turbulent substructure occurring inside the large vortices in our simulation runs p180 and p360. \cite{Lesur+Papaloizou2009} showed a similar effect for vortices with aspect ratio $\chi > 8$ influenced by elliptic instability. The larger vortices in our simulations have aspect ratios $ 8 < \chi < 11 $ and fit  well with the predictions. We do not observe the predicted decay of these vortices by the elliptic instability, similar to the reports of \cite{Lesur+Papaloizou2009}. In how far this an effect of an inactive elliptic instability (EI), maybe suppressed by low resolution, or an overlap with the VSI is open for future investigations. At $\chi = 10$ the minimum growth time of the elliptic
instability should be about 17.6 Orbits. But maybe the vortices receive additional driving
from absorbing smaller vortices (as observed for Jupiter's Red Spot), which so far counteract destruction from the EI.

The occurrence of the EI also naturally explains the fast destruction of the small vortices we observe in all simulation runs. These vortices have $\chi < 4$ and are strongly influenced by the fast growing modes of the elliptic instability reported for this size regime.

\section{Summary \& Conclusions}
\label{sec:summary}

We have performed full 3D hydrodynamical simulations of protoplanetary disks undergoing Vertical Shear Instability. We used four different azimuthal extents ranging between $\pi/2$ and $2 \pi$ to asses the influence of non-axisymmetry on the development of the instability.

We summarize our main findings as follows:
\begin{itemize}
\item We find the Vertical Shear Instability to be capable of seeding vortices with large ($\chi > 8$) aspect ratios using the disk parameters $p=-\frac{2}{3}$, $q=-1$ and $\frac{H}{R}=0.1$. This has to our knowledge not been reported for simulations of this instability. The vortices we observe are long lived (lifetimes larger than 500 local orbits) and can aid in the growth of planetesimals in protoplanetary disks via particle trapping as proposed by \cite{Barge+Sommeria1995}.
 
\item We find the angular momentum transport in VSI to be sufficiently efficient and in agreement with latest assumptions for protoplanetary disks and constraints on planetesimal formation therein \citep{Drazkowska+Alibert2017}. Interrestingly, despite most angular momentum is transported outward at large $z$, the radial mass flux is also outward at large $z$ but inward close to the midplane. This is due to additional \revisedvtwo{strong} vertical transport of angular momentum in agreement with \citep{Stoll+Kley2017}. \revisedvtwo{This leads to transport of angular momentum away from the midplane into upper layers, where it is then transported outwards.}

\item As a direct consequence, depending on the height $z$ above the midplane, the ration of $v_\mathrm{rms}(z)$ to $\sqrt{\alpha}$ ranges between 2.0 and 3.5. This height depending ratio has to be taken into consideration when inferring disk $\alpha$ values from rms-velocity measurements.

\item The choice of the size of the simulation domain, especially of $\phi_\mathrm{max}$ has a significant impact on the outcome of the simulation. We find for $\phi_\mathrm{max} < 90$ no indication for larger, long lived vortices forming. Furthermore we find a systematic increase of $\alpha$-values and rms-velocities with decreasing $\phi_\mathrm{max}$.
\end{itemize} 

Our findings for the VSI on smaller scales are therefore consistent with the results reported by \cite{Nelson+2013},\cite{Stoll+Kley2014,Stoll+Kley2016},\cite{Richard+2016} and \cite{Flock+2017} for the low azimuthal extent cases, while identifying their limitations on more global disk scales. We propose future work on instabilities in protoplanetary disks include global $360^\circ$ simulations to identify possible low m effects suppressed in current simulations of disks sections.

Furthermore, we find resolution as an important factor, both for the development of the RWI and the EI in global disk simulations. We suggest the lower radial resolution used in \cite{Stoll+Kley2014} as a possibility why they did not find vortex formation in their work. Although they use different parameters in their work ($p=-1.5$, $q=-1$ and $\frac{H}{R}=0.05$), we do not find significant differences to their work in preliminary results of a parameter study we conduct. The results of this study will be the subject of a follow up publication. We also propose further high resolution simulations should be performed to investigate the influence of the EI on the longevity of the vortices generated.

\section*{Acknowledgements}
The authors indebted to discussions with Willy Kley.
We also would like to thank the other members of the theory of planet and star formation group at the Max-Planck-Institute for Astronomy in Heidelberg for their useful advice, patience and help with this project. 

This research has been supported by the Deutsche Forschungsgemeinschaft Schwerpunktprogramm (DFG SPP) 1833 "Building a Habitable Earth" under contract KL1469/13-1 "Der Ursprung des Baumaterials der Erde: Woher stammen die Planetesimale und die Pebbles? Numerische Modellierung der Akkretionsphase der Erde." This research was supported by the Munich Institute for Astro- and Particle Physics (MIAPP) of the DFG cluster of excellence "Origin and Structure of the Universe and in part  at KITP Santa Barbara by the National Science Foundation under Grant No. NSF PHY11-25915. 
The authors gratefully acknowledge the Gauss Centre for Supercomputing (GCS) for providing computing time for a GCS Large-Scale Project (additional time through the John von Neumann Institute for Computing (NIC)) on the GCS share of the supercomputer JUQUEEN \citep{Stephan:202326} at J\"ulich Supercomputing Centre (JSC) and the GCS Supercomputer HAZEL HEN at H\"ochstleistungsrechenzentrum Stuttgart (www.hlrs.de). GCS is the alliance of the three national supercomputing centres HLRS (Universit\"at Stuttgart), JSC (Forschungszentrum J\"ulich), and LRZ (Bayerische Akademie der Wissenschaften), funded by the German Federal Ministry of Education and Research (BMBF) and the German State Ministries for Research of Baden-W\"urttemberg (MWK), Bayern (StMWFK) and Nordrhein-Westfalen (MIWF). Additional simulations were performed on the THEO and ISAAC clusters owned by the MPIA and the HYDRA and DRACO clusters of the Max-Planck-Society, all four hosted at the Max-Planck Computing and Data Facility in Garching (Germany).

\appendix
\section{2D resolution study}
\begin{figure}
\centering
\includegraphics[width=\columnwidth]{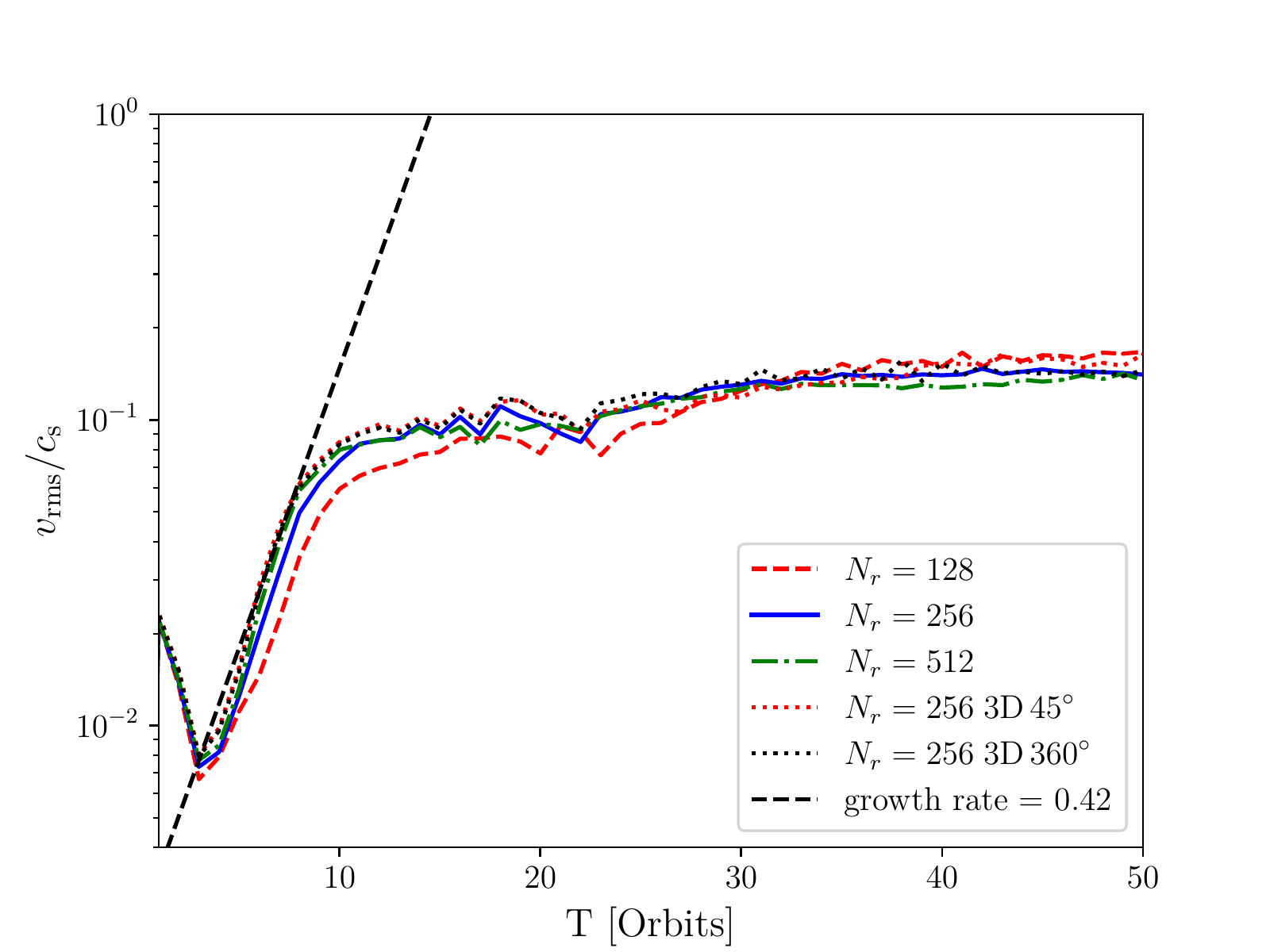}
\caption{Rms velocities for the 2D resolution study in radial direction. We find convergence for the domain sizes above $N_r=256$. For comparison, we also plot the values from the 3D model presented in figure \ref{fig:vrmsvT}.}
\label{fig:vrms2Drad}
\end{figure}

\begin{figure}
\centering
\includegraphics[width=\columnwidth]{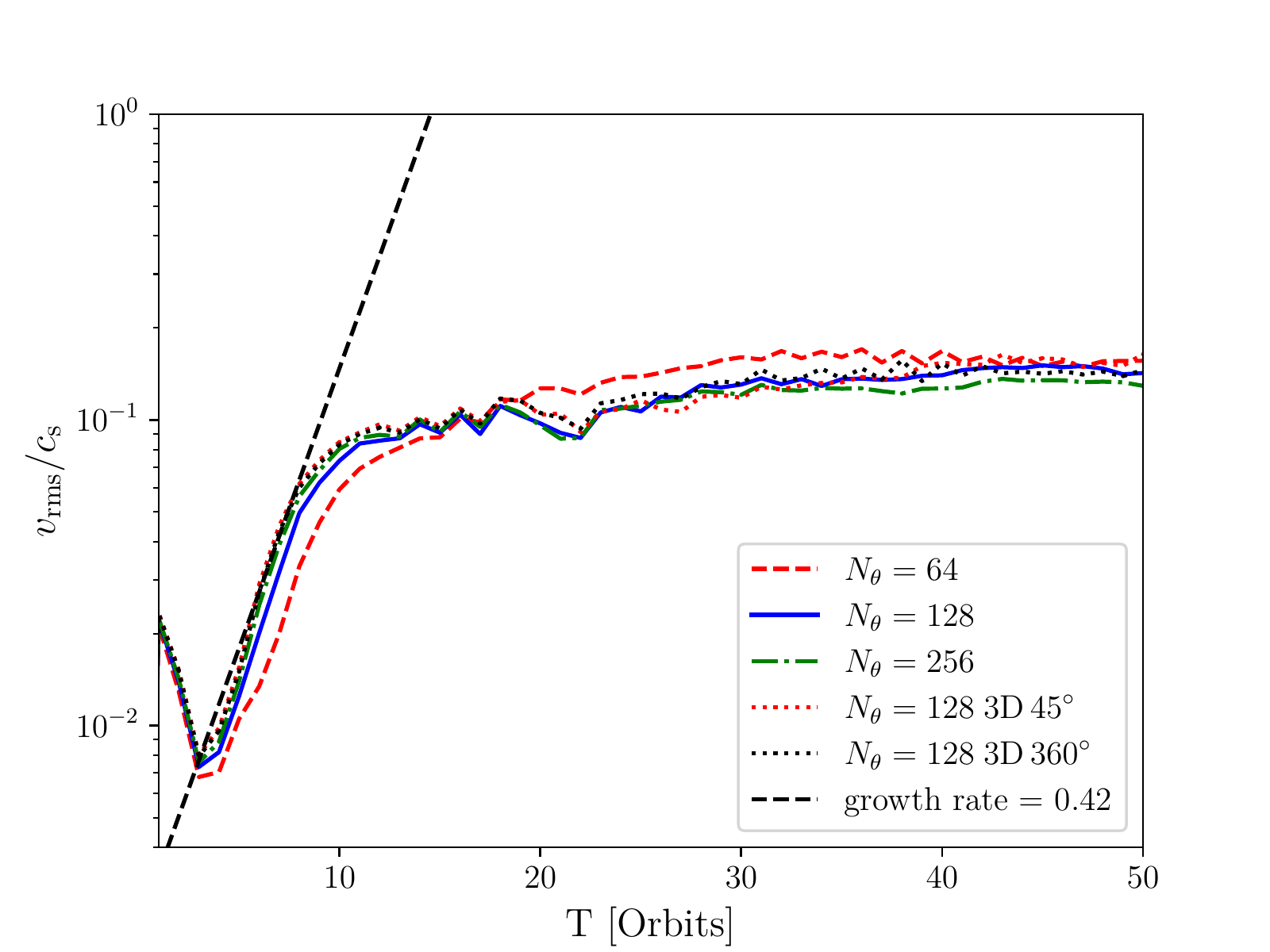}
\caption{Rms velocities for the 2D resolution study in vertical direction. We find convergence for the domain sizes larger or equal $N_\theta=128$. For comparison, we again also plot the values from the 3D model presented in figure \ref{fig:vrmsvT}.}
\label{fig:vrms2Dmer}
\end{figure}

We performed a resolution study in 2D to confirm we resolve the instability in the radial \revised{and meridional} direction. We conducted 3 runs with radial domain sizes of $N_r$ = 128, 256 and 512 \revised{and $N_\theta$ = 128 and 3 additional runs with $N_r$ = 256 and $N_\theta$ = 64, 128 and 256}. In figures \ref{fig:vrms2Drad} \revised{and \ref{fig:vrms2Dmer}} we show the RMS-velocities for the initial growth phase.

We find convergence for radial domain sizes $N_r \geq 256$ \revised{and $N_\theta \geq 128$}, proving the values we chose for the 3D simulations to be sufficient to capture the relevant physics in the early growth phases of our simulations. We also show the slope for a growth rate of $\Gamma = 0.42$ per orbit as a guidance value (black dashed curve).\revised{The growth rates found for the 2D simulations are in good agreement with the results presented above for the 3D case, supporting the validity of our choice of simulation domain parameters}.

\label{lastpage}

\bibliography{references}

\end{document}